\documentclass[11pt]{article}
\usepackage[utf8]{inputenc}
\usepackage[T1]{fontenc}
\usepackage[sort&compress,numbers,super]{natbib}
\usepackage[version=3]{mhchem}
\usepackage{pgfplots}
\usepackage{subcaption}
\usepackage{tikz}
\usepgfplotslibrary{groupplots}
\pgfplotsset{compat=1.17}
\usepackage{footmisc}
\usepackage{physics}
\usepackage{xspace}
\usepackage[separate-uncertainty=true,multi-part-units=single]{siunitx}
\usepackage{authblk}
\usepackage[margin=1 in]{geometry}
\author[1]{Ryker Fish}
\author[2]{Adam Carter}
\author[3]{Pablo Diez-Silva}
\author[3]{Rafael Delgado-Buscalioni\thanks{Email: \texttt{rdbuam@gmail.com}}}
\author[3]{Raul P. Pelaez\thanks{Email: \texttt{raulppelaez@gmail.com}}}
\author[1]{Brennan Sprinkle\thanks{Email: \texttt{bsprinkl@mines.edu}}}

\affil[1]{Department of Applied Mathematics and Statistics, Colorado School of Mines, Golden, CO, USA}
\affil[2]{CNRS, Sorbonne Université, Physicochimie des Electrolytes et Nanosystèmes Interfaciaux, Paris, France}
\affil[3]{Department of Theoretical Condensed Matter Physics, Universidad Autónoma de Madrid, Madrid, Spain}

\usepackage{float}\tolerance=1000
\usepackage{bm}
\usepackage{amsmath, amssymb}
\usepackage{graphicx}
\usepackage{caption}
\usepackage{hyperref}
\usepackage{cleveref}
\usepackage{setspace}
\usepackage{color}
\usepackage{enumerate}
\usepackage{listings}
\usepackage[frozencache,cachedir=minted-cache]{minted}
\setminted{fontsize=\normalsize}
\usepackage{tcolorbox}
\tcbuselibrary{minted, skins, breakable}

\newtcblisting{CodeBox}[2][]{
  listing engine=minted,
  colback=white,
  colframe=black,
  boxrule=1pt,
  sharp corners,
  breakable,
  listing only,
  minted language=#2,
  minted options={
    breaklines=true,
    breakanywhere=true,
    fontsize=\small,
    autogobble=true,
    tabsize=2
  },
  left=5pt,
  right=5pt,
  top=5pt,
  bottom=5pt,
  before skip=10pt,
  after skip=10pt,
  before upper={\vspace*{-0.5\baselineskip}\singlespacing},
  #1 
}

\renewcommand{\vec}[1]{\bm{#1}}
\newcommand{\tens}[1]{\bm{\mathcal{#1}}}
\newcommand{\oper}[1]{\mathcal{#1}}
\newcommand{\grandm}{\bm{\mathcal{M}}}
\newcommand{\kT}{k_B T}
\newcommand{\drift}[1]{\kT \partial_{\vec{X}} \cdot \grandm^{#1}}

\newcommand{\grandr}{\bm{\mathcal{R}}}
\newcommand{\deltaR}{\Delta \grandr}
\newcommand{\exactR}{\grandr^{\text{exact}}}
\newcommand{\mobR}{\grandr^{\text{mob}}}
\newcommand{\lubmob}{\overline{\grandm}}

\newcommand{\noise}{\mathcal{W}}
\DeclareMathOperator*{\argmax}{\arg\!\max}

\hypersetup{pdfborder=0 0 0}
\date{\today}
\hypersetup{
 pdftitle={libMobility: A Python library for hydrodynamics at the Smoluchowski level},
 pdfkeywords={},
 pdfsubject={},
 pdflang={English}}

\title{libMobility: A Python library for hydrodynamics at the Smoluchowski level}

\newcommand{\libM}{\texttt{libMobility}\xspace}
\newcommand{\DPStokes}{\texttt{DPStokes}\xspace}
\newcommand{\NBody}{\texttt{NBody}\xspace}
\newcommand{\PSE}{\texttt{PSE}\xspace}

\begin{document}
\maketitle

\begin{abstract}
  Effective hydrodynamic modeling is crucial for accurately predicting fluid-particle interactions in diverse fields such as biophysics and materials science. Developing and implementing hydrodynamic algorithms is challenging due to the complexity of fluid dynamics, necessitating efficient management of large-scale computations and sophisticated boundary conditions. Furthermore, adapting these algorithms for use on massively parallel architectures like GPUs adds an additional layer of complexity.
  This paper presents the \libM software library, which offers a suite of CUDA-enabled solvers for simulating hydrodynamic interactions in particulate systems at the Rotne-Prager-Yamakawa (RPY) level. The library facilitates precise simulations of particle displacements influenced by external forces and torques, including both the deterministic and stochastic components. Notable features of \libM include its ability to handle linear and angular displacements, thermal fluctuations, and various domain geometries effectively. With an interface in Python, \libM provides comprehensive tools for researchers in computational fluid dynamics and related fields to simulate particle mobility efficiently. This article details the technical architecture, functionality, and wide-ranging applications of \libM.
  \libM is available at \url{https://github.com/stochasticHydroTools/libMobility}.

\end{abstract}
\section{Introduction}
\label{sec:intro}
Understanding the dynamical properties of fluid-particle systems at the micron-scale is an essential problem in soft matter physics, microbiology, nanotechnology, and a litany of other fields. The long-ranged, many-body nature of hydrodynamic interactions and the importance of Brownian motion at this scale pose difficulties for computational models\cite{Delong_BD_without_GF_2014,SpectralRPY,Delmotte_Usabiaga_multiblob_review_2025} but are crucial components for the quantitative predictions of colloidal suspensions \cite{EleanorCountoscope2024, RollersLubrication,Bossis_Brady_Rheology_1989,Gao2025}, cell-scale biology \cite{ParkKimBacteriaSwimming2022,SubcellularHI_Review}, microfluidics\cite{Microfluidics_Review,MultiscaleMicrofluidics_Review}, polymeric fluids\cite{PolymerTumbling_PRL,PolymerChainMD_Dunweg,SemiflexFluct,ActinCLsRheology}, etc. Effective hydrodynamic modeling enables researchers and engineers to characterize fundamental physical mechanisms and optimize the design of complex materials and systems.

Despite its importance and maturity, the landscape of computational hydrodynamics software remains fragmented. Existing solutions vary greatly in their capabilities, usability, availability, and level of ongoing maintenance. Many specialized solvers are scattered across distinct repositories, sometimes without publicly available code or accessible documentation, which significantly impedes reproducibility and collaborative advancement in the field. Additionally, algorithms designed for particular geometries or boundary conditions often require vastly different computational approaches (see \cite{SpectralRPY, StokesDP, ForceCoupling_Fluctuations, Ermak_McCammon_1978}), presenting challenges when adapting or comparing results across different settings. The lack of a unified and intuitive interface further complicates the effective use of hydrodynamic modeling tools by a broad community of users.

Since the advent of Graphics Processing Units (GPUs) as generic computing hardware, especially fostered by the release of CUDA\cite{cuda2008}, the scientific community has put a tremendous effort into developing algorithms specifically tailored for GPUs \cite{ForceCoupling_Fluctuations, Fast_FCM_Keaveny_2024, StokesDP, PoissonDP, UAMMD_Raul_2025, SpectralSD, OpenMM8_2024, anderson2020hoomd, LAMMPS2022, kohnkeGromacsFMM2020}. GPUs can dramatically reduce computational times for large-scale simulations, making previously intractable hydrodynamic problems accessible. Leveraging the CUDA programming model enables efficient use of GPU capabilities, allowing for substantial performance improvements over traditional CPU-based implementations.

In this work, we introduce \libM, a GPU-accelerated library with a Python interface specifically designed to simulate hydrodynamic interactions at the Smoluchowski level, i.e. at the level of overdamped Langevin dynamics \cite{roux_brownian_1992, Ermak_McCammon_1978, VACF_Langevin}.

\libM addresses the existing challenges by providing a coherent, modular interface through which researchers can seamlessly switch between various hydrodynamic algorithms tailored for different geometries. By exploiting CUDA and building upon GPU-optimized numerical modules from existing libraries such as UAMMD\cite{UAMMD_Raul_2025}, \libM achieves significant computational efficiency while maintaining ease of use through its Python front-end.

The remainder of this paper is structured as follows: Section \ref{sec:theory} provides the theoretical foundations underlying the hydrodynamic models implemented in \libM. Section \ref{sec:library} details the software architecture, installation instructions, usage examples, and available GPU-enabled solvers. Section \ref{sec:validation} presents extensive validation tests demonstrating the accuracy and robustness of \libM. Section \ref{sec:examples} showcases illustrative examples including passive and active colloidal suspensions, rheological measurements, and electro-osmotic flows. Finally, Section \ref{sec:conclusions} summarizes the key contributions and outlines future directions for ongoing development and community engagement.
\section{Theory}
\label{sec:theory}
We are interested in computing the displacements of a collection of microscopic, spherical particles submerged in a fluctuating, zero--Reynolds fluid experiencing conservative, external forces. The fluctuating Stokes equations govern the dynamics of the fluid according to
\begin{align} \label{eq:stoke1}
  \eta\nabla^2\vec{v} = \nabla p - \tilde{\vec{f}} - \nabla\cdot\boldsymbol{\mathcal{Z}}, \hspace{0.5cm} \nabla\cdot\vec{v} = 0
\end{align}
Where $\vec{v}$ is the fluid velocity, $p$ is the pressure, $\eta$ is the viscosity of the fluid, $\tilde{\vec{f}}$ is a force density acting on the fluid, and $\boldsymbol{\mathcal{Z}}$ is a tensorial fluctuating stress with mean zero and covariance given by:
\begin{equation}
  \langle \mathcal{Z}_{ik}(\vec{x},t)\mathcal{Z}_{jm}(\vec{x'},t')\rangle = 2\kT\eta(\delta_{ij}\delta_{km} + \delta_{im}\delta_{kj})\delta(\vec{x}-\vec{x'})\delta(t-t'),
\end{equation}
which is chosen to ensure that the system satisfies the fluctuation-dissipation theorem\cite{Landau:Fluid}.

The force density $\tilde{\vec{f}}$ represents forces felt by the submerged particles and has two components,
\begin{equation}
  \tilde{\vec{f}} = \vec{f} + \vec{f}_{th}.
  \label{eq:fpcoup}
\end{equation}
The first term, $\vec{f}$, comes from the external forces ($\vec{F}$) and torques ($\vec{T}$) acting on the particles. This is expressed in an Immersed Boundary (IB) framework via
\begin{equation}
  \vec{f}(\vec{x}) = \sum_{i=1}^N \left[\vec{F}_i\delta_F(\vec{x}-\vec{x}_i) + \frac{1}{2} \nabla\times\vec{T}_i\delta_T(\vec{x}-\vec{x}_i)\right],
\end{equation}
where the force density $\vec{f}$ at the point $\vec{x}$ has contributions from $N$ particles with labels $i\in[1,N]$ located at positions $\vec{x}_i$. This IB formulation of Brownian dynamics is described in more detail by Delong \textit{et al.} \cite{Delong_BD_without_GF_2014} The monopolar (force) and dipolar (torque) contributions are spread to the fluid over a finite volume following the IBM mediated by regularized envelope functions $\delta_F$ and $\delta_T$ centered at particles. Different models use different regularization functions to cater to different boundary conditions, particle geometries, or computational efficiency needs. Examples include Gaussian kernels\cite{ForceCoupling_Fluctuations} or Peskin kernels\cite{IBM_PeskinReview}.

On the other hand, the so-called thermal drift term in \cref{eq:fpcoup}, represented as ${\vec{f}}_{th}$, is a forcing of thermal origin necessary to ensure that the system satisfies detailed balance. \cite{SELM,roux_brownian_1992} Following the discussion in Sec. II.A of Delong \textit{et al.}, \cite{Delong_BD_without_GF_2014} this term can be written as
\begin{equation}
  \vec{f}_{th} = -\kT\vec{\partial}_{\vec{X}}\sum_{i=1}^{N}\delta(\vec{x}-\vec{x}_{i}),
  \label{eq:tdparts}
\end{equation}
where $\vec{\partial}_{\vec{X}}$ is the gradient operator with respect to the particle positions.

We impose the kinetic constraint on the particles,
\begin{equation}
  \begin{aligned}
    \vec{u}_i = \frac{d\vec{X}_i}{dt}           & = \int \vec{v}(\vec{x},t)\delta_F(\vec{x}-\vec{x}_i)d\vec{x}                         \\
    \vec{\omega}_i = \frac{d\vec{\theta}_i}{dt} & = \frac{1}{2} \int (\nabla\times\vec{v}(\vec{x}))\delta_T(\vec{x}-\vec{x}_i)d\vec{x}
  \end{aligned}
  \label{eq:kinetic_constraint}
\end{equation}
where $\vec{X}_i$ and $\vec{\theta}_i$ are the position and orientation of the $i$th particle, respectively. $\vec{u}_i$ and $\vec{\omega}_i$ are the translational and angular velocities of the $i$th particle, respectively.

We may write the Stokes equations \eqref{eq:stoke1} more compactly as:
\begin{equation}
  \vec{v} = \oper{L}(\tilde{\vec{f}} - \nabla\cdot\boldsymbol{\mathcal{Z}})
\end{equation}
where $\oper{L}$ is the Stokes solution operator, aka the Green's function of the Stokes equations. The operator $\oper{L}$ is a linear operator that maps the force density to the velocity field. The particular shape of this operator depends on the geometry of the system, but can be written in a general form as
\begin{equation}
  \oper{L} = -\eta^{-1}\nabla^{-2}\left(\mathbb{I}-\nabla\nabla^{-2}\nabla \cdot \right).
\end{equation}

It is possible to eliminate the fluid from the description entirely by writing \cref{eq:kinetic_constraint} in terms of the Green's function of the Stokes equations. By defining the $6N\times 6N$ (where $N$ is the number of particles in the system) grand mobility matrix $\grandm$, as
\begin{equation}
  \label{eq:block_mobility}
  \grandm = \begin{bmatrix}
    \grandm_{\vec{u}\vec{F}}      & \grandm_{\vec{u}\vec{T}}      \\
    \grandm_{\vec{\omega}\vec{F}} & \grandm_{\vec{\omega}\vec{T}}
  \end{bmatrix}
\end{equation}
where, e.g.,
\begin{equation}
  \label{eq:stokestomob}
  \grandm_{{\vec{u}\vec{F}}}(\vec{x}_i, \vec{x}_j) = \iint \delta_{F}(\vec{x}-\vec{x}_i)\oper{L}(\vec{x},\vec{x'})\delta_{F}(\vec{x'}-\vec{x}_j)d\vec{x}d\vec{x'}
\end{equation}
we can succinctly express the stochastic evolution of particle positions and orientations.
The equations of motion for the particles now take the form:
\begin{equation}
  \begin{bmatrix}
    d\vec{X} \\
    d\vec{\theta}
  \end{bmatrix} = \grandm\begin{bmatrix}
    \vec{F} \\
    \vec{T}
  \end{bmatrix} dt + \sqrt{2\kT\grandm}d\vec{\noise} + \kT\vec{\partial}_{\vec{X}}\cdot\grandm dt,
  \label{eq:bdhi}
\end{equation}
where $d\vec{X}$ and $d\vec{\theta}$ are the linear and angular displacements of the particles, $\vec{F}$ and $\vec{T}$ are the forces and torques acting on the particles, and $d\vec{\noise}$ is a vector of independent Gaussian random variables with zero mean and variance $dt$. The second term in \cref{eq:bdhi} arises from the thermal fluctuations in the fluid. Its magnitude is determined by the fluctuation-dissipation theorem, which states
\begin{equation}
  \label{eq:fdt}
  \left\langle\left(2\kT\grandm\right )^{1/2}d\vec{\noise}\left[\left(2\kT\grandm\right )^{1/2}d\vec{\noise}\right]^T\right\rangle = 2\kT dt \grandm.
\end{equation}
Finally, the third term in \cref{eq:bdhi} represents the thermal drift described in \cref{eq:tdparts}. Note that since our particles are radially symmetric, the resulting mobility is independent of the particles' orientation, and thus the thermal drift term is only non-zero for the translational degrees of freedom. It is worth noting that in situations where the mobility matrix is translationally invariant and isotropic (such as in a triply periodic environment or in a bulk fluid), the thermal drift term vanishes entirely. This is not the case, however, in confined geometries such as slit channels.
In geometries with confining boundaries the mobility becomes a function of distance to the boundary and so the divergence $\partial_{\vec{X}} \cdot \grandm$ is non-zero in general. The divergence is also known to be non-zero in grid-based immersed boundary methods because a finite-width kernel cannot provide perfect translational invariance. In this case, the thermal drift term must be included to correct for the lack of translational invariance even in the fully periodic case \cite{Delong_BD_without_GF_2014}.

The generic form in \cref{eq:bdhi} can accommodate different geometries, boundary conditions, and models by modifying the solution operator of the Stokes equations and the regularization functions. Even in the few cases where the analytical expression of this solution operator is known, direct application of \cref{eq:bdhi} is often not feasible due to the size of the mobility matrix. Further, accounting for thermal fluctuations requires the calculation of the matrix square root of $\grandm$, which is an $O(N^3)$ operation if a naive Cholesky factorization is used. Thus, historically the community has developed specialized solvers for each geometry, boundary condition, and even hardware to subvert these high computational costs.
\libM attempts to unify the myriad of specialized solvers available in the literature by providing a common interface for the calculation of each term in \cref{eq:bdhi}. Importantly, the library is designed to be modular, allowing for the addition of new solvers and geometries in the future.
\section{The libMobility library}
\label{sec:library}
\subsection{Getting started}
At the time of writing, \libM is designed primarily as a GPU-focused library, leveraging CUDA extensively and thus requiring an NVIDIA GPU for execution. While pre-built packages currently support only Linux and will work on the average high-performance computing system, \libM can also be compiled from source on Windows.
\libM is open-source software distributed under the permissive MIT license. Its source code is publicly available at \url{https://github.com/stochasticHydroTools/libMobility}.

Installation of \libM can be conveniently performed using the Conda package and environment manager\cite{conda_software} via the \texttt{conda-forge} channel \cite{conda_forge_community_2015_4774217}:

\begin{CodeBox}[]{bash}
  conda install -c conda-forge libmobility
\end{CodeBox}

Once installed, \libM can be readily imported into Python scripts and environments. Comprehensive documentation, including detailed usage instructions and theoretical background, is available at \url{https://libmobility.readthedocs.io}.

\subsection{The libMobility interface}
The \libM library provides a unified interface across its GPU-accelerated solvers in the form of interchangeable Python classes with methods to compute each of the terms in \cref{eq:bdhi}. The library is designed to be modular, allowing for the addition of new solvers and geometries in the future. In particular, each solver presents the following interface:

\begin{itemize}
  \item Constructor: Creates the solver specifying the periodicity in each direction. Each solver can only accommodate a specific subset of the available geometries, so the user should check the documentation for each solver to see which are supported. Currently, available options are: \texttt{periodic}, \texttt{open}, \texttt{single\_wall} and \texttt{two\_walls}.
  \item \texttt{setParameters}: This method is used to set domain-specific parameters. For example, this allows a user to specify the location of the confining walls for solvers with the \texttt{single\_wall} and \texttt{two\_walls} geometries.
  \item \texttt{initialize}: This method is used to set parameters shared by all solvers, such as the viscosity of the fluid and hydrodynamic radius of the particles.
  \item \texttt{setPositions}: Updates the solver with the current particle positions.
  \item \texttt{Mdot}: Given one or both of forces and torques acting on the particles, this method computes the resulting deterministic displacements $\grandm \vec{F}$.
  \item \texttt{sqrtMdotW}: Compute the thermal fluctuations of the particles, $\displaystyle \grandm^{1/2}\vec{\mathcal{W}}$.
  \item \texttt{divM}: Compute the thermal drift term for the particles, $\partial_{\vec{X}} \cdot \grandm$.
  \item \texttt{LangevinVelocities}: Compute all three terms in \cref{eq:bdhi}. This is cheaper as a combined step in some solvers- see below. Defaults to an Euler-Maruyama step for solvers without a specific implementation.
\end{itemize}
Some algorithms can compute multiple terms of \cref{eq:bdhi} simultaneously at a lower computational cost than computing each term separately. The \texttt{LangevinVelocities} method takes advantage of this when possible. In the current version of \libM, only the \PSE solver has a specialized implementation that computes the deterministic and stochastic terms more cheaply.

Below is an example of how to use the \libM library to compute the displacements of a collection of particles in a domain with a bottom wall. The example uses the \NBody solver which implements RPY kernels, as discussed in \Cref{sec:available_solvers}. The example begins by importing the necessary libraries and creating an instance of the \NBody solver with a bottom wall geometry. The solver is then initialized with the necessary parameters, including the height of the wall in $z$, fluid viscosity, and hydrodynamic radius of the particles. We have additionally included the optional \texttt{includeAngular} parameter so that the solver will accept torques and return angular velocities. Without it, only forces are accepted and linear velocities are returned. This parameter defaults to false since the solvers, in general, are more efficient if angular velocities are not needed for an application. The positions of the particles are then set using the \texttt{setPositions} method. The forces and torques acting on the particles are then defined and passed to the \texttt{Mdot} method to compute the deterministic displacements. The thermal fluctuations are computed using the \texttt{sqrtMdotW} method, and finally, the thermal drift term is computed using the \texttt{divM} method. The total displacements are then computed by summing all terms. In practice, these terms should have coefficients relating to the energy $\kT$ and timestep of a simulation. Note that all methods return two values that correspond to linear and angular displacements.
\begin{CodeBox}[]{python}
  from libMobility import NBody
  import numpy as np

  solver = NBody("open", "open", "single_wall")
  solver.setParameters(wallHeight=0.0)
  solver.initialize(viscosity=1.0, hydrodynamicRadius=1.0, includeAngular=True)

  pos = np.random.rand(100, 3) * 10.0
  forces = np.random.rand(100, 3)
  torques = np.random.rand(100, 3)

  solver.setPositions(pos)
  mf, mt = solver.Mdot(forces=forces, torques=torques)
  dw_f, dw_t = solver.sqrtMdotW()
  drift_f, drift_t = solver.divM()

  dX = mf + dw_f + drift_f
  dtheta = mt + dw_t + drift_t
\end{CodeBox}

When providing \libM with a \texttt{numpy} array, the library will automatically copy the data to the GPU. The library will accept any array object that complies with the buffer protocol (via DLPack \url{https://github.com/dmlc/dlpack}), which includes \texttt{numpy} arrays, \texttt{cupy} arrays and even \texttt{pytorch} or \texttt{jax} tensors. Passing an array already in GPU memory will avoid the overhead of copying the data to and from the GPU. In general, the library will respond with a tensor of the same type as the input one that have been passed to the different methods. This means that sending the forces as a \texttt{numpy} array, will result in a CPU-GPU transfer for the input, and a GPU-CPU transfer for the output.

\subsection{Thermal fluctuations}
Some solvers provide a specialized way to compute thermal fluctuations efficiently for a particular geometry. One example is the triply periodic \PSE solver, a spectral method that incorporates fluctuations in a natural and inexpensive way in Fourier space. When a solver does not implement its own version of the fluctuations (i.e. it only presents a method to apply $\grandm$), \libM will employ the Lanczos algorithm \cite{SquareRootKrylov}, a Krylov subspace decomposition method, to compute the square root of the mobility matrix via an iterative application of the mobility operator. This method is independent of the geometry and the details of the solver as long as the mobility matrix is symmetric positive definite. Currently, Lanczos is used for the \DPStokes and \NBody solvers within \libM. Previous studies have investigated the convergence of the Lanczos algorithm for these solvers \cite{StokesDP, MagneticRollers}. The convergence of the Lanczos algorithm can be improved in some situations via the use of a preconditioner\cite{SquareRootPreconditioning}. Although the current \libM interface could accommodate such functionality, the current release does not make use of a preconditioner.

\subsection{Thermal drift}
\label{sec:thermal_drift}

In a similar spirit as with the thermal fluctuations, \libM provides a way to compute the thermal drift term for solvers that do not provide a specialized implementation. The default implementation simply sets the thermal drift to zero since this is a common case in purely open or periodic geometries. For solvers where this is not the case, \libM offers an implementation of the Random Finite Difference (RFD) method \cite{Delong_BD_without_GF_2014}, which approximates the thermal drift as
\begin{equation}
  \label{eq:rfd}
  \vec{\partial}_{\vec{X}}\cdot\grandm = \partial_j \mathcal{M}_{ij}(\vec{X}) = \frac{1}{\delta} \bigg\langle \bigg( \mathcal{M}_{ij} \bigg(X_k + \frac{\delta}{2} W_k \bigg) W_j - \mathcal{M}_{ij} \bigg(X_k - \frac{\delta}{2} W_k \bigg) W_j \bigg) \bigg\rangle + O(\delta^2),
\end{equation}
where $\vec{W}$ is a vector of independent Gaussian random variables with mean zero and variance one, and $\delta$ is a small parameter that is chosen to be small enough to ensure convergence of the method without incurring in numerical accuracy issues.

Note that since our particles are radially symmetric the mobility does not the depend on the orientation of the particles, i.e. $\partial_{\vec{\theta}} \cdot \grandm = 0$, and only the divergence of the mobility with respect to the positions is non-zero\cite{roux_brownian_1992}.

\subsection{Available solvers}
\label{sec:available_solvers}

\libM presently bundles three production‐ready hydrodynamic solvers that are exposed through the common Python interface described above, as well as an additional example solver that demonstrates how to add a new solver to \libM and can be used to run a simulation with only single-body hydrodynamic interactions. Each solver targets a different set of available geometries and has a different balance of computational cost so that practitioners can select the most adequate algorithm for their application. A concise overview is given below, and we point users to the official documentation (hosted at \url{https://libmobility.readthedocs.io}) for further details. Table~\ref{tab:solver_summary} summarizes the key capabilities of each solver. Since all solvers are either periodic or open in $x$ and $y$, we omit that information and only list the supported boundaries in $z$. We include asymptotic scaling information for the number of particles $N$ and, where applicable, the size of the domain, $L$.

\paragraph{SelfMobility}
The \texttt{SelfMobility} module neglects inter–particle hydrodynamic interactions and applies the Stokes drag tensor \(\grandm= (6\pi\eta a)^{-1}\vec{I}\) independently to every particle.  Because no long‑range flow needs to be resolved, the method operates in \(O(N)\) time and memory and is useful for verification tests or for dilute suspensions where hydrodynamic coupling is negligible. Only fully \emph{open} boundaries are accepted in all three directions. Importantly, this solver provides a simple template for how a new solver could be added to \libM.

\vspace{-5pt}
\paragraph{Positively Split Ewald (PSE).}
For triply–periodic domains, the preferred work‑horse is the GPU implementation of the Positively Split Ewald algorithm of Fiore \textit{et al.}\cite{SpectralRPY} \PSE solves the Stokes equations through a real/reciprocal–space decomposition of the Green's function and samples the Gaussian fluctuations directly in Fourier space. The real part of the decomposition is evaluated in a purely Lagrangian manner (using only the neighborhoods of each particle), while the reciprocal space is computed by translating the particles into a grid (Eulerian description). Thus, PSE is a hybrid Eulerian-Lagrangian method.

In practice, the solver scales nearly linearly with system size (and exactly linear with the number of particles) and yields the thermal noise at the cost of one forward and one inverse FFT.
All three directions \emph{must} be declared \texttt{periodic}. This solver also exposes a non-Ewald-split version of the solver, which is functionally equivalent to the fluctuating Force Coupling Method by Keaveny\cite{ForceCoupling_Fluctuations}. In the current version of \libM, the \PSE solver can only be used with forces.

\vspace{-5pt}
\paragraph{RPY kernels (NBody)}
The \NBody solver is a Greens function-based solver, providing a brute‑force \(O(N^{2})\) evaluation of pairwise RPY mobilities. This solver excels at small systems (less than 50K particles) or when the spatial extent of the domain is too large to be simulated quickly using a grid-based method.
The transverse directions \(x\) and \(y\) are restricted to \texttt{open}. The \(z\) direction can be chosen as \texttt{open} (free-space RPY) or \texttt{single\_wall}, in which case the Swan-Brady kernel \cite{StokesianDynamics_Wall} is used. Notably, the Swan-Brady kernel does not produce a symmetric positive definite (SPD) mobility matrix when particles overlap the wall. To address this, we have included the damping matrix approach in Appendix A from the authors of \cite{MagneticRollers} that ensures the mobility of a particle smoothly goes to zero as a particle approaches the wall in a way that maintains the SPD property of the mobility matrix.

\vspace{-5pt}
\paragraph{Doubly Periodic Stokes (DPStokes)}
For doubly periodic domains, the \DPStokes solver offers an \emph{asymptotically linear} algorithm based on the recent spectral method of Hashemi \textit{et al.} \cite{StokesDP} The solver mandates \texttt{periodic} boundaries in \(x\) and \(y\); along \(z\), the user may select \texttt{open}, \texttt{single\_wall}, or \texttt{two\_walls}. For the \texttt{two\_walls} geometry, particles are required to stay between the walls. However, note that a simulation domain must be prescribed for all geometries in \DPStokes. This is because \DPStokes is a grid-based method with a finite extent in $z$, and particles are not allowed to leave the grid. This height must be tuned to balance efficiency with physics as the solver will produce an error when a particle exits this artificial grid boundary.

\vspace{-5pt}
\subsection{When to use each solver}
Often, the geometry of the application dictates the solver one will need to employ. When simulating particles near no-slip boundaries, \NBody and \DPStokes offer the \texttt{single\_wall} geometry with one planar wall in $z$. In this geometry, \DPStokes is the more efficient solver for applications with a large number of particles as it has a finite domain size via periodic boundary conditions in the $xy$-plane. The \NBody solver is better for applications with few particles, unbounded domains, or particles with small radii compared to the simulation domain. This last point is due to \DPStokes being a grid-based method, so gridding a large domain and using a small particle radius is computationally expensive. See \cref{sec:diffusion} and \cref{sec:timings} for further discussion about choosing between the \NBody and \DPStokes solvers. \DPStokes is the only solver which offers the \texttt{two\_walls} geometry, where a suspension is confined between two planar walls in $z$.

\begin{figure}
  \centering
  \includegraphics[width=1.0\linewidth]{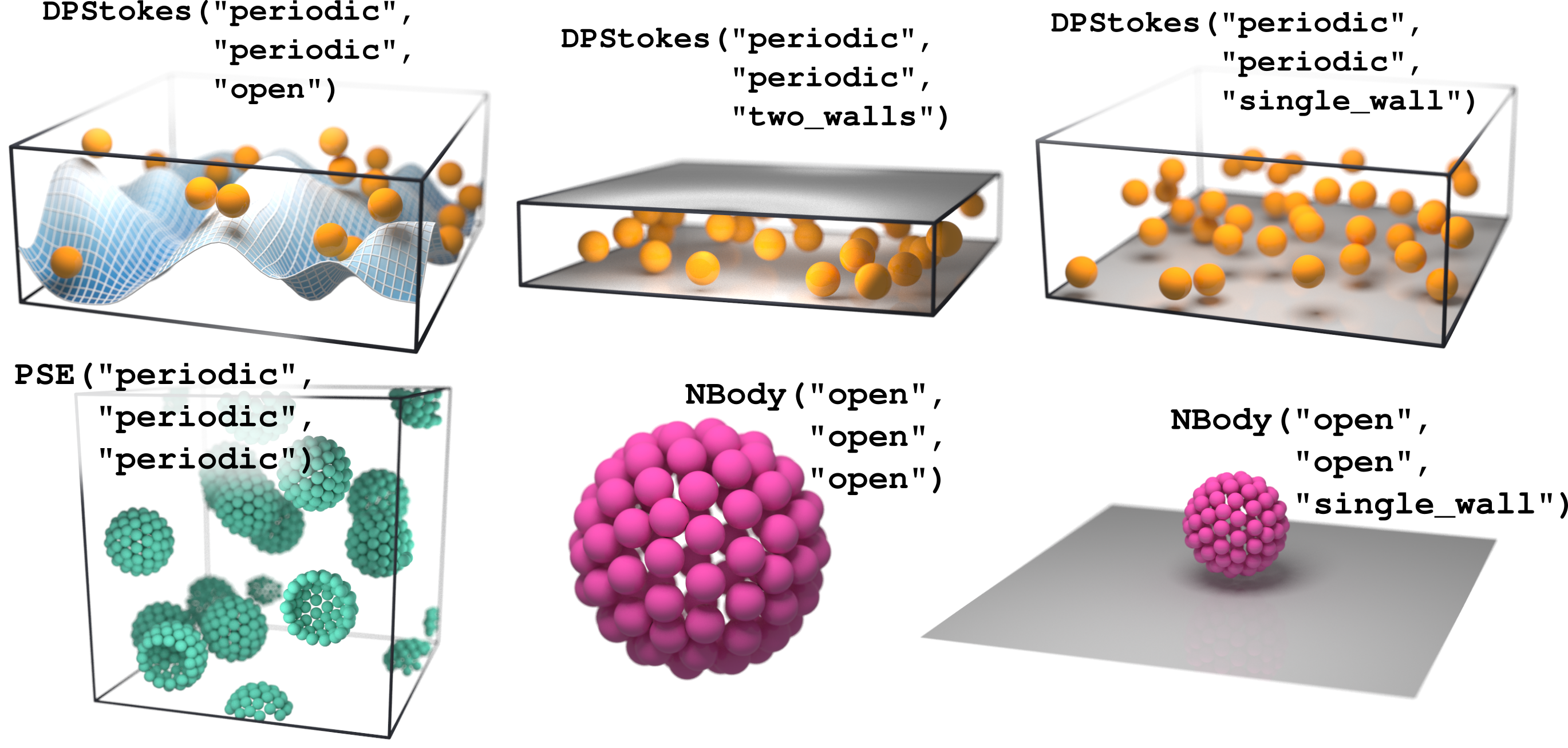}
  \caption{Illustration of domain geometries supported by the release version of \libM. Domains encased in a unit box indicates periodic boundary conditions. A code snippet for initialization of the solver with the shown geometry is included above each image.}
  \label{fig:solver_illustration}
\end{figure}

For suspensions far from no-slip boundaries, \NBody offers fully open domains, \PSE supports triply periodic domains, and \DPStokes can be set to use open conditions in $z$ with periodic boundary conditions in the $xy$-plane. \NBody is best for a small number of particles or necessary when a suspension in an unbounded domain. \PSE is much more efficient than \NBody for a large number of particles and can be used for simulations at fixed packing fractions due to the finite (periodic) domain. \PSE additionally has functionality to apply a shear to a suspension, but does not currently support torques/angular velocities. The open mode in \DPStokes is intended for quasi-2D suspensions such as freespace monolayers or planar membranes, see \cref{fig:solver_illustration} for an illustration. Finally, \texttt{SelfMobility} is offered as a template for how a new solver could be added to \libM as a module, although it could also be used as a drop-in replacement for a dry Brownian dynamics simulation. The capabilities of the solvers are summarized in \cref{tab:solver_summary}.

\begin{table}[H]
  \centering
  \caption{Summary of solver capabilities shipped with the release version of \libM. $\dag$ indicates the solver is periodic in the $xy$--plane.}
  \label{tab:solver_summary}
  \begin{tabular}{lcccc}
    \hline
    Solver         & Geometry $(z)$    & Cost (N)     & Cost (L)   & Typical use‐cases          \\
    \hline
    SelfMobility   & open              & \(O(N)\)     & --         & Dry Brownian dynamics      \\
    PSE$\dag$      & periodic          & \(O(N)\)     & $O(L^3)^*$ & Bulk colloidal suspensions \\
    NBody          & open/single\_wall & \(O(N^{2})\) & --         & Large domains              \\
    DPStokes$\dag$ & open/1–2 walls    & \(O(N)\)     & $O(L^3)$   & Dense monolayers           \\
    \hline
  \end{tabular}
  \parbox{0.9\linewidth}{\footnotesize $^*$ \PSE can see greatly improved scaling over what is reported in the table e.g for cases involving a small, constant number of particles, the complexity is largely unaffected by increasing the domain size.}
\end{table}

\section{Validation}
\label{sec:validation}
\libM includes a comprehensive test suite designed to rigorously verify both the correctness and numerical stability of its solvers. The tests are organized into three primary categories: deterministic hydrodynamic mobility, fluctuation-dissipation, and thermal drift consistency. All test can be run using \texttt{pytest}\cite{pytest}. Overall, these tests guarantee that each solver computes deterministic and stochastic particle displacements correctly. Additionally, several tests assessing the stability of the library API are present, checking error conditions and interface guarantees to increase usability. Validation tests are run before every new version of \libM is released to ensure the library continues to provide the expected behavior as new features or optimizations are added.

\paragraph{Deterministic hydrodynamic mobility}
Deterministic tests are conducted by directly comparing computed particle displacements under known force distributions against established analytical or numerically benchmarked solutions. For instance, pairwise hydrodynamic interactions computed by the \NBody solver are verified against exact analytical solutions for the Rotne-Prager-Yamakawa (RPY) and Swan-Brady kernels. Additionally, periodic solvers such as \PSE and \DPStokes are validated by comparing against existing spectral and Ewald-type numerical results available in literature.

\paragraph{Fluctuation-dissipation theorem}
The fluctuation-dissipation theorem (FDT) ensures that thermal fluctuations generated by the solver accurately reflect the hydrodynamic interactions described by the mobility operator, as given by \cref{eq:fdt}. To test this, \libM evaluates the statistical consistency between deterministic mobility matrices and stochastic displacements. For each solver, the computed thermal fluctuations are compared against the theoretical covariance provided by the mobility operator. Specifically, we test that
\begin{equation}
  \vec{Z} = \grandm^{-1/2}_{\text{SVD}} \grandm^{1/2} \vec{W} \sim N(\vec{0}, \vec{I}),
\end{equation}
where the subscript SVD denotes we compute that inverse square root exactly using the singular value decomposition of the full mobility matrix $\grandm$. This transformation is done for convenience, so that we may perform  statistical tests on each component $\vec{Z}$ independently. Agreement is evaluated statistically over multiple realizations via a Kolmogorov-Smirnov test\cite{KSTest}, confirming the solver correctly implements the relationship between deterministic mobility and stochastic fluctuations as mandated by the FDT. Sanity checks, such as the mobility matrix being positive definite, are also performed.

\paragraph{Thermal drift}
Validation of the thermal drift term is performed by checking that the average of many RFDs converges to the divergence of the mobility computed using a deterministic finite difference. The thermal drift term can be computed deterministically as
\begin{equation}
  (\partial_{\vec{q}} \cdot \grandm(\vec{q}))^{det} = \frac{1}{\delta} \sum_{j=1}^{M} \left[\grandm(\vec{q}^+) \vec{e}_j - \grandm(\vec{q}^-)\vec{e}_j \right],
\end{equation}
where $\vec{e}_j$ is a unit vector with only the $j$-th entry non-zero, $\vec{q}^{\pm} = \vec{q} \pm \frac{\delta}{2} \vec{e}_j$, and $M = 6N$ when both forces and torques are included and $M=3N$ for only forces. This is used to validate the RFD implementation of thermal drift as
\begin{equation}
  \big\Vert (\partial_{\vec{q}} \cdot \grandm)^{det} - \langle (\partial_{\vec{q}} \cdot \grandm)^{RFD} \rangle \big\Vert_\infty < \epsilon,
\end{equation} where the average is taken over enough realizations of the RFD to converge within a specified error tolerance $\epsilon$.

\section{Examples}
\label{sec:examples}
The following examples are chosen to demonstrate the speed and flexibility of \libM. We consider four applications that highlight the advantages of different solvers, as well as interfacing \libM with other software libraries for multi-physics applications. Code to recreate each example can be found at \url{https://github.com/stochasticHydroTools/libmobility-paper-examples}.

\subsection{Active colloids (NBody)}
\label{sec:rollers}
We use a minimal model for a suspension of torque-driven particles above a bottom wall, following Usabiaga \textit{et al.} \cite{MagneticRollers}. The particles are embedded with a hematite cube that imparts a small magnetic moment to the particle which can then can be rotated with an oscillating magnetic field. The presence of the bottom wall breaks symmetry of the problem and the hydrodynamic interaction with the wall turn the rotation of the particles into translational motion. It is known that the strong collective flows created by the rotation causes large groups of particles to move faster than any individual particle could and break off into groups known as ``critters''. \cite{Rollers_NaturePhys, TwoLines_Rollers} Unless stated otherwise, all parameters and the simulation procedure come from Usabiaga \textit{et al.} \cite{MagneticRollers}

\begin{figure}[h]
  \centering
  \includegraphics[width=0.98\linewidth]{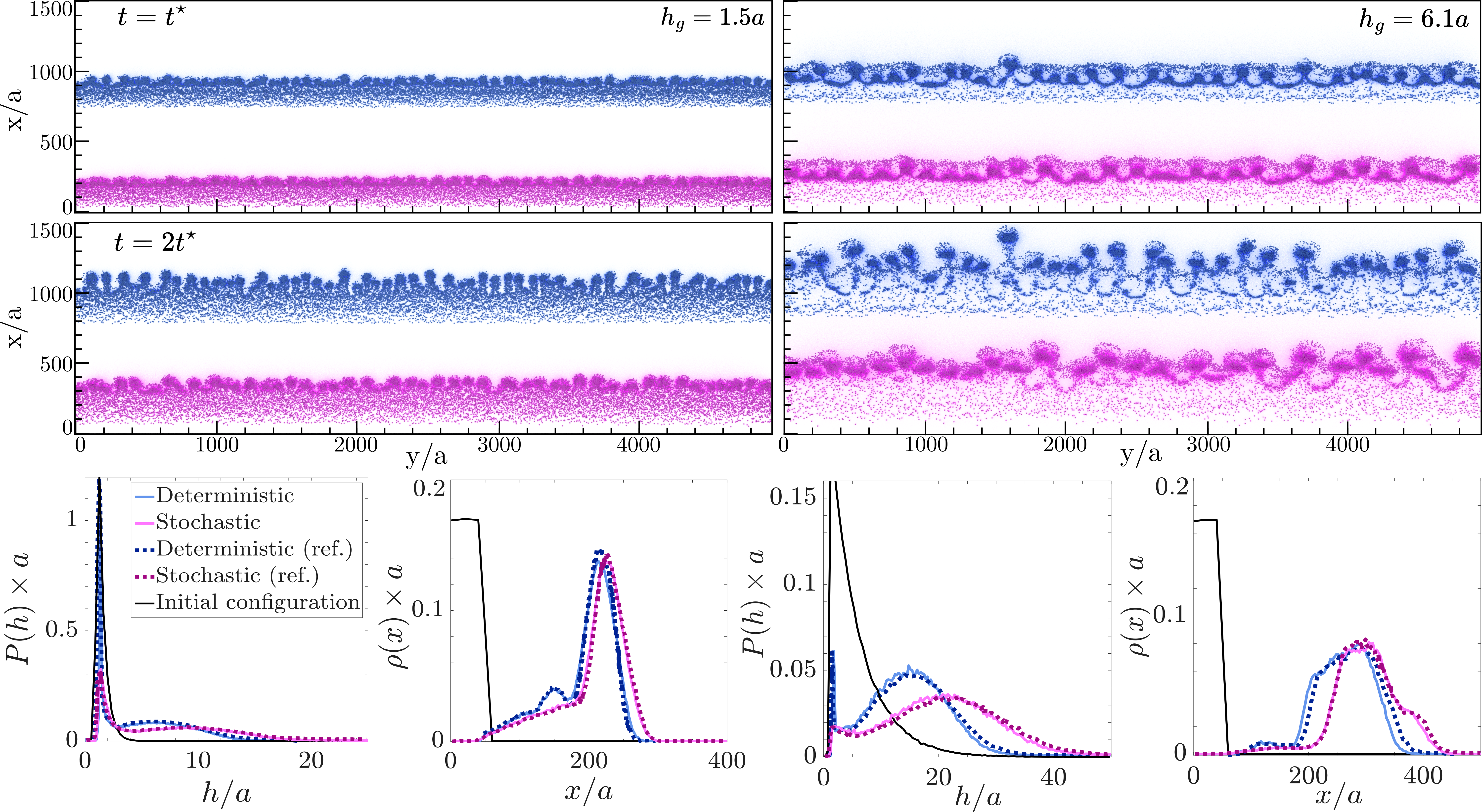}
  \caption{Comparison of gravitational heights $h_g=1.5a$ (left) and $h_g=6.1a$ (right) for deterministic (blue) and stochastic (pink) simulations. Top panels: top-down snapshots of simulations at multiples of the characteristic time $t^\star$. Deterministic and stochastic simulations are independent and offset for visual clarity. Bottom panels: probability distribution of particle heights at $t=t^\star$ (left) and distribution of particle positions at $t=t^\star$ (right). The distributions at $t=0 s$ are shown in black. Reference data\cite{MagneticRollers} is shown in darker colors and with dashed lines.}
  \label{fig:rollers}
\end{figure}

We simulate \cref{eq:bdhi} using the \NBody solver with the \texttt{single\_wall} geometry from \libM both deterministically and stochastically ($k_BT = 0$ aJ and $k_BT = 4.11 \times 10^{-3}$ aJ, respectively) for $N=2^{15}=32768$ particles of radius $a=0.656 \mu m$. The average height of the particles is given by the gravitational height $h_g = a + k_BT / mg$ where $m$ is the excess mass of the particle and $g$ denotes gravity. Simulations are performed for different values of $h_g = (1.5, 6.1)a$ by changing the mass of the particles. The initial configuration of particles was created by gridding a domain in the $x$-$y$ plane with dimensions $60a \times 4915a$ with grid cells of size $2a$ so that no particles overlap, then randomly selecting cells to place colloids in. The initial height of each particle was then sampled from the equilibrium Gibbs-Boltzmann distribution $P(z) \propto \exp(-mgz/k_BT)$. This initialization procedure differs from that Usabiaga \textit{et al.}, however we conclude that the final results shown in \cref{fig:rollers} are insensitive to the initialization procedure.

During simulation, a soft pairwise repulsive steric potential is included between particles and between particles and the wall to prevent excessive overlap. The form of the potential is
\begin{equation}
  \label{eq:soft_potential}
  U(r) = U_0 \begin{cases}
    1 + \frac{d-r}{b},   & r < d     \\
    \exp(\frac{d-r}{b}), & r \geq d,
  \end{cases}
\end{equation}
where $r$ is the distance between the center of a particle and another particle or to the wall. We take $d=2a$ for particle-particle interactions and $d=a$ for particle-wall interactions. Following Usabiaga \textit{et al.}, we use $U_0 = 4k_BT$ and $b=0.1a$. Taking this relatively large value for the interaction range $b$ results in a ``soft'' potential that will eliminate overlap but do so slowly enough to prevent restricting the timestep to be unnecessarily small. In addition to the potential above, we also include a hard-core steric potential only between particles and the wall. The potential (equation 5, Varga \textit{et al.} \cite{ColloidalGelation_Swan}) is designed to eliminate overlap between two particles hydrodynamically interacting with the free space RPY kernel in exactly one timestep. Since the potential depends on the mechanism for energy dissipation, the presence of a wall in the simulation means overlaps between a particle and the wall can persist for more than one timestep. Regardless, we found that including the hard-core potential substantially improved convergence of the Lanczos algorithm for computing the noise in the stochastic simulations due to the decreased number of particles overlapping the wall.

To discretize \cref{eq:bdhi}, we use the stochastic Adams-Bashforth scheme,
\begin{align}
  \vec{X}^{n+1} = \vec{X}^n & + \Delta t \Biggr\{ \left( \frac{3}{2} \grandm^n \vec{F}^n - \frac{1}{2} \grandm^{n-1} \vec{F}^{n-1} \right) \nonumber \\
                            & + \left(\frac{3}{2}\drift{n} - \frac{1}{2} \drift{n-1} \right) \Biggl\} \nonumber                                      \\
                            & + \sqrt{2\kT\Delta t} {(\grandm^n)}^{1/2} \vec{W}^n,
\end{align}
where superscripts denote the timestep, $\vec{F}$ represents a combined vector of forces and torques, and $\vec{W}$ has mean zero and variance one, as in \cref{eq:rfd}. Note the slight difference from Usabiaga \textit{et al.} in that we also include the thermal drift term in the deterministic part of the time stepping scheme as it has non-zero mean.

To compare the structure of the instability across different gravitational heights and between simulations with and without stochasticity, we compute a characteristic time $t^\star$ for the formation of the fingering instability. The characteristic time is defined as
\begin{equation}
  \label{eq:roller_characteristic_time}
  t^\star = \argmax_t \int_{k_{\min}}^{k_{\max}} \vert \hat{n}(k_y, t) \vert^2 dk_y,
\end{equation}
where $\hat{n}(k_y, t)$ is the Fourier transform of the 1D number density $n(y;t)$, or equivalently the Fourier transform of the 2D number density $n(x,y;t)$ at wavenumber $k_x=0$ (eq. 10, Usabiaga \textit{et al.}\cite{MagneticRollers}). Only the furthest 70\% of particles are used to compute $\hat{n}$ to neglect particles that have fallen behind the wavefront. We found the need to adjust the lower wavenumber cutoff $k_{\min}$ in the integral to avoid picking up on signals caused by wavelengths larger than the average size of the critters. We use $k_{\min}=(0.15, 0.1) \mu m^{-1}$ for $h_g=(1.5, 6.1)a$, respectively, and $k_{\max}=0.25 \mu m^{-1}$ for both. The top panels of \cref{fig:rollers} show snapshots for both gravitational heights at $t=t^{\star}, 2t^\star$. The simulation for $h_g=1.5a$ has critters with a smaller lateral spread in the $y$ direction than the simulation for $h_g=6.1a$, which corresponds to the need for a larger cutoff for $k_{\min}$ in \cref{eq:roller_characteristic_time} when $h_g=1.5a$. \Cref{tab:characteristic_times} shows the characteristic times for all four cases averaged over multiple simulation runs. Identical to Usabiaga \textit{et al.}, we see it takes $1.2$x and $1.4$x longer for the instability to form in the stochastic case when $h_g=1.5a$ and $h_g=6.1a$, respectively.

\begin{table}[H]
  \centering
  \caption{Mean and std. of characteristic times from \cref{eq:roller_characteristic_time}, (n=4)}
  \label{tab:characteristic_times}
  \begin{tabular}{lcc}
                      & $h_g=1.5a$     & $h_g=6.1a$     \\
    \hline
    Deterministic (s) & $9.1 \pm 0.7$  & $28.5 \pm 1.0$ \\
    Stochastic (s)    & $11.2 \pm 0.7$ & $39.9 \pm 2.2$ \\
    \hline
  \end{tabular}
\end{table}

To evaluate the structure of the instability, we compute the distribution of particle heights $P(h)$ and the distribution of particle positions along the direction of travel $\rho(x)$ at $t^\star$. The lower panels of \cref{fig:rollers} show we are accurately able to reproduce the distributions from the reference data using \libM for all cases, including the smoother distributions that come from correctly including the stochastic dynamics.

\subsection{Passive colloids (NBody)}
\label{sec:diffusion}
\newcommand{\Dself}{D_\mathrm{self}}

Another application of \libM is to simulate passive colloidal systems to investigate diffusion of individual particles, as well as collective dynamics of large groups of particles. Here, we use \libM to obtain large-scale simulations of particles diffusing above a wall and in free space.

In a sufficiently dilute suspension of colloids, each colloid will diffuse with self diffusion coefficient $\Dself$. In the 2D case, this quantity can be computed from the mean squared displacement (MSD) of the particle as
\begin{equation}
  \Dself = \lim_{t\rightarrow 0} \frac{1}{4t} \left\langle [ \vec{r}(t_0 + t) - \vec{r}(t_0) ]^2 \right\rangle
  \label{eq:msd}
\end{equation} where $\vec{r}(t)$ is the position at time $t$, and $\langle \cdot \rangle$ means averaging over all particles and all initial times. The self diffusion coefficient can also be found from the intermediate scattering function (ISF) $F(k, t)$, sometimes known as the dynamic structure factor. The ISF is the correlation of the Fourier transformed particle density field $\rho(\vec{k}, t)$: $F(k, t) =\langle \rho(\vec{k}, t_0 + t) \rho^*(\vec{k}, t_0) \rangle / N$ where $N$ is the number of particles. Evaluated at $t=0$, the ISF gives the (static) structure factor $F(k, 0) = S(k)$, which describes the relations between the positions of the particles in a suspension but encodes no dynamic information.

The ISF governs the decay of density fluctuations. Fluctuations of a wavelength $k$ decay as
\begin{equation}
  \frac{F(k, t)}{F(k, 0)} = \exp\left\{-k^2D(k)t\right\}
  \label{eq:fkt exp decay}
\end{equation}
where we have defined a wavevector-dependent diffusion coefficient $D(k)$. In the large wavevector (short length scale) limit, $D(k)$ converges to the self diffusion coefficient: $D(k\rightarrow\infty) = \Dself$ \cite{dhont_introduction_nodate}, while the small wavevector (large length scale) limit defines the collective diffusion coefficient. In a hard-sphere colloidal suspension without hydrodynamic interactions between particles, $D(k)$ is given by
\begin{equation}
  D(k) = \frac{\Dself}{S(k)}
  \label{eq:Dk}
\end{equation}
and converges as $k\rightarrow 0$, giving a well-defined collective diffusion coefficient \cite{dhont_introduction_nodate}. In a 3D suspension of colloids with hydrodynamic interactions, $D(k)$ is also known to converge in the small wavevector limit \cite{segre_scaling_1996, banchio_short-_2018}. However, results differ in a quasi-2D suspension where colloids are restriction to a plane but the solvent extends in 3D. In this setting, hydrodynamic interactions cause the collective diffusion coefficient to diverge as $k^{-1}$ \cite{bleibel_hydrodynamic_2014, panzuela_collective_2017}, and a similar effect has been also observed in the short-time collective diffusion of membrane lipids\cite{PanzuelaDelgadoBuscalioni2018}.
However, the behavior of the collective diffusion coefficient in the small-$k$ limit for a quasi-2D suspension above a wall has been until now unreported, but \libM allows us to access the large length- and time-scales needed to probe this regime.

\subsubsection{Simulation method}
Here we focus on simulating purely diffusing particles above a bottom wall. We require two modifications from that used in \cref{sec:rollers} to achieve results that agree quantitatively with experiments. First, theoretical predictions and experimental systems use particles that are hard spheres, i.e. they cannot overlap \cite{AdamCollectiveDiffusion2025, EleanorCountoscope2024}. To model this in simulation, we use the same form of the potential in \cref{eq:soft_potential} but decrease the interaction length parameter to $b=2a\delta / \ln(10)$ so that the potential decays to $0.01 U_0$ at a separation distance of $r=d(1+\delta)$. Following Sprinkle \textit{et al.}\cite{RollersLubrication}, we use $\delta=10^{-2}$. Second, lubrication corrections are needed to improve the accuracy of near-field hydrodynamics. Since \libM uses regularized kernels to approximate the mobility matrix, near-field hydrodynamics are poorly resolved. This results in particles moving too quickly when near a surface and produces a mean squared displacement higher than seen in experiments. Including lubrication corrections causes the diffusing particles to move more slowly when near the bottom wall and the resulting MSD agrees closely with experiments. Here, we only simulate dilute suspensions with in-plane packing fraction $\phi = 0.11$ and the particles tend to be well-separated, thus inter-particle lubrication corrections are neglected for simplicity but would be required to simulate denser suspensions. Our method closely follows that of Sprinkle \textit{et al.} \cite{RollersLubrication} but stripped of any inter-particle corrections.

We briefly describe lubrication corrections in general, then describe our simple version. While far-field interactions can be approximated as pairwise using a mobility formulation, near-field interactions can be approximated pairwise using a resistance formulation, e.g. by using the resistance matrix $\grandr = \grandm^{-1}$. Using $\grandm_{\text{pair}}$ as the mobility matrix for a pair of nearby particles (or a particle and a wall), lubrication corrections pairwise correct the resistance matrix by subtracting off the near-field component $\mobR = \left( \grandm_{\text{pair}} \right)^{-1}$ and adding on the ``exact'' pair interaction $\exactR$, which may come from an asymptotic approximation or a pre-tabulation of a highly-resolved numerical method; a thorough summary of available analytical formulas is available in \cite{Townsend_Lub}. Using $\deltaR = \exactR - \mobR$, the lubrication-corrected mobility matrix $\overline{\grandm}$ can be stated as\cite{RollersLubrication}
\begin{equation}
  \label{eq:lub_mobility}
  \overline{\grandm} = \grandm \left[\boldsymbol{I} + \deltaR \grandm \right]^{-1}.
\end{equation}
Instead of building this matrix explicitly, the action of $\lubmob$ on a vector can be computed by solving the system
\begin{equation}
  \left[\boldsymbol{I} + \grandm \deltaR \right] \mathbf{x} = \grandm \mathbf{F}
\end{equation}
using an iterative Krylov solver.

Lubrication corrections are most often applied to $\grandm$ when it has been constructed by truncating the multipole expansion after the dipole terms, but they can be applied in the same manner when using different far-field approximations for $\grandm$ \cite{FastStokesianDynamics_2019, RollersLubrication}. To obtain a method that is fast specifically for simulating dilute passive particles, our reduced method only uses the $\grandm_{\vec{u}\vec{F}}$ block of \cref{eq:block_mobility} and thus only applies a correction $\deltaR$ that is 3x3 and diagonal, corresponding only the interactions between a single particle and the bottom wall. To make lubrication corrections fast, values for $\mobR$ are pre-tabulated by directly inverting the 3x3 mobility matrix for a particle approaching a wall at many values of the dimensionless height $\displaystyle \epsilon_h = \frac{\boldsymbol{q} \cdot \hat{\boldsymbol{z}}}{a} - 1$ and then interpolated during the simulation. Values for $\exactR$ are computed from the asymptotic expansions of Goldman \textit{et al.} \cite{Goldman1967} for $\epsilon_h < \left(0.3, 0.15\right)$  for the horizontal and vertical coefficients, respectively, and are pre-tabulated (then later interpolated) using a multiblob with 2562-blobs for larger $\epsilon_h$; see \cref{sec:rheology} for a brief description and relevant references for the multiblob method. To simulate particle dynamics, we use Algorithm 1 from the authors of \cite{RollersLubrication} which only requires a method to compute deterministic displacements, thermal noise, and the drift term, as well as a method to apply $\deltaR$. We use the \NBody solver in \libM for the mobility matrix, noise, and drift term above a bottom wall, and use the construction for $\deltaR$ described above. The \NBody solver requires the boundaries to be open in the $x$ and $y$ directions, so in order to maintain a bounded domain for the monolayer to diffuse in we add a potential in the $x$/$y$ directions of the form
\begin{equation}
  U(r_i) = \frac{k_B T}{a^2}
  \begin{cases}
    r_i{}^2     & r_i < 0           \\
    0           & 0 \leq r_i \geq L \\
    (r_i - L)^2 & r_i > L
  \end{cases}
\end{equation}
for $i \in \{x, y\}$ and where $L$ is the side length of the domain. Note that we could have alternatively used the doubly periodic \DPStokes solver to maintain a finite domain size, but we found that the \NBody solver is $\sim 10\%$ faster for the domain size and particle density we consider. For this example, we empirically found the efficiency break-even between \DPStokes and \NBody to be approximately $L/a=1800$, with \DPStokes eventually becoming more efficient for a larger number of particles; see \cref{sec:timings} for more details. This efficiency crossover depends on both the domain size, packing fraction $\phi$, and the graphics card used to run the simulation, and should therefore be re-evaluated for different applications. Fortunately, this is straightforward to do as \libM facilitates easy exploration of different solvers and geometries.

\subsubsection{Results}
We simulate a diffusing monolayer at $\phi=0.11$ at domain size $L = \SI{2560}{\micro\meter}$. The timestep used is $\SI{0.125}{\second}$ and 8 hours of simulated data are collected.
We calculate the short-time self diffusion coefficient from \cref{eq:msd} and obtain a value of $\Dself=\SI{0.042}{\micro\meter\squared\per\second}$, compatible with the value of $\SI{0.043\pm0.001}{\micro\meter\squared\per\second}$ measured in experiments \cite{AdamCollectiveDiffusion2025}.

The intermediate scattering function $F(k, t)$ is calculated from the resulting particle positions via the direct method:
$
  F(k, t) = \left\langle \sum_\mu \sum_\nu \exp{-i\vec{k} \cdot [\vec{r_\mu}(t) - \vec{r_\nu}(0)]^2} \right\rangle
$
over a grid of $k$ points spaced by $\flatfrac{2\pi}{L}$. Note that we only use the $x$ and $y$ (in-plane) coordinates when computing this quantity. For more details, see supplementary information of ref. \cite{AdamCollectiveDiffusion2025}, section 1.3. To extract $D(k)$, we invert \cref{eq:fkt exp decay}, however, one must choose a time $t$ to make the inversion at. In the large-$k$ regime, the ISF decays quickly and so we perform the inversion at $t=\SI{1}{\second}$. At longer times, the ISF would have already reached its noise floor.
Conversely, in the small-$k$ regime, the ISF barely begins to decay at short times which leads to large uncertainties in the data. To avoid this, we perform the inversion at $t=\SI{1024}{\second}$. Although the self diffusion coefficient varies with time, the collective diffusion coefficient is expected to be the same at short and long times for our system of hard-sphere particles \cite{dhont_introduction_nodate, pusey_liquids_1991}.

\begin{figure}
  \centering
  \includegraphics[width=1.0\linewidth]{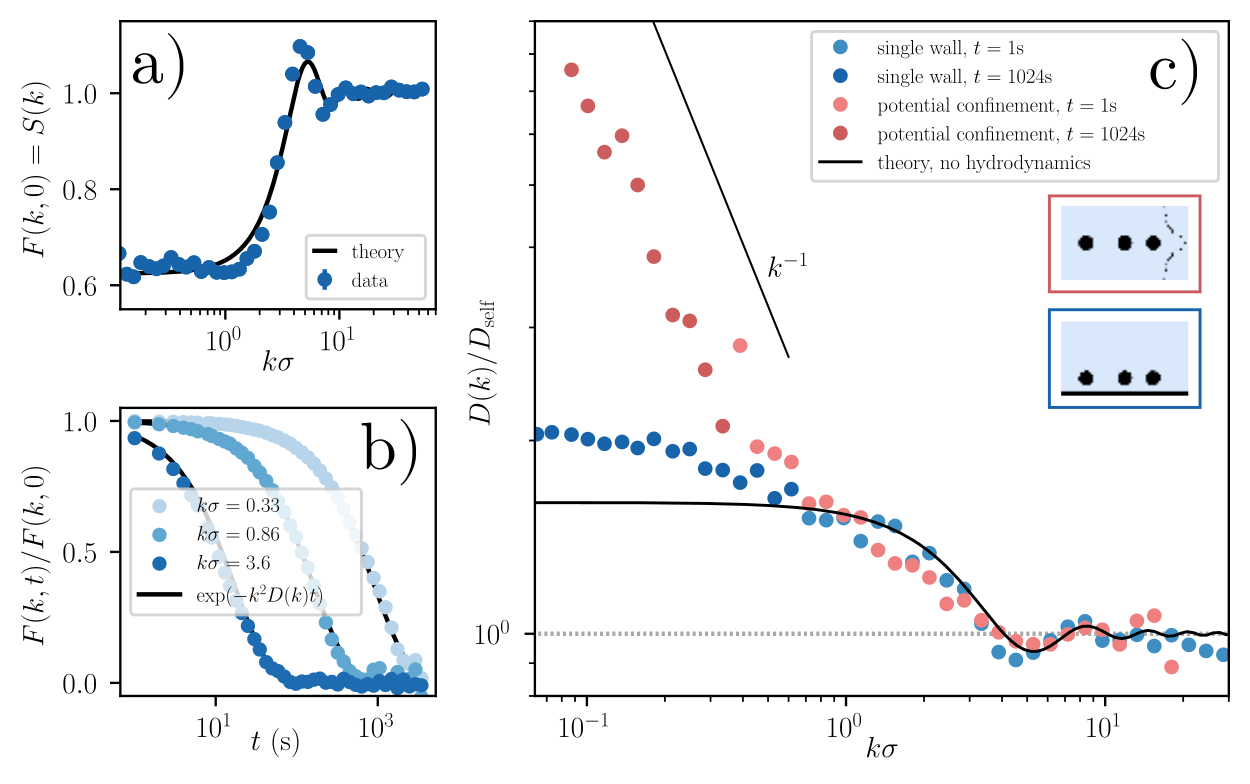}
  \caption{a) structure factor $S(k)$ for the simulation of a monolayer above a wall. The theoretical curve is from Thorneywork \textit{et al.}\cite{thorneywork_structure_2018}. b) example decays of $f(k, t)$, from the same simulation, along with fitted curves. c) $D(k)$ for the monolayer suspended in free space (red), and above a wall (blue), showing clearly, the lack of divergence, compared to in free space. The black theory curve is \cref{eq:Dk}.
  }
  \label{fig:collective diffusion}
\end{figure}

In order to verify our simulation and analysis method, we also perform simulations for the case of a quasi-2D monolayer embedded in a fully open 3D geometry. The monolayer is kept in place by a quadratic potential of width $\sigma$ of the form $U(z) = (z - z_0)^2 k_s/2$, where $k_s = k_B T / \sigma^2$. We use $\sigma=a/2$ to keep colloids confined in a narrow region. Shown in red in \cref{fig:collective diffusion}c, we clearly recover the $k^{-1}$ divergence expected in this situation \cite{bleibel_hydrodynamic_2014,panzuela_collective_2017}.

In \cref{fig:collective diffusion}c we present $D(k)$ for the monolayer suspended above a wall, where the confinement is provided purely by gravity and electrostatic repulsion from the wall, using the same physical parameters as from Mackay \textit{et al.}\cite{EleanorCountoscope2024}. We see, for the first time, substantial evidence of a plateau in $D(k)$ in the small $k$ limit. Hydrodynamic interactions have caused the collective diffusion coefficient to increase by a factor of approximately $1.3$ compared to the theoretical collective diffusion coefficient without hydrodynamic interactions, but have not lead to a divergence.
The extremely large length- and time-scales that characterize collective diffusion above a wall is the reason why this collective diffusion coefficient has been hard to measure, in contrast to a quasi-2D monolayer embedded in open 3D space. Due to the slower diffusive dynamics above a wall, the time scale of the decay of a density fluctuation with $k\sigma=10^{-1}$ (which corresponds to a length scale of $\sim \SI{200}{\micro\meter}$) is approximately 3 hours. \libM allows us to perform simulations of micron-scale colloids with domain sizes on the order of millimeters for long enough times to accurately resolve these slow dynamics.

\subsection{Rheology (PSE)}
\label{sec:rheology}
\newcommand{\bu}{\boldsymbol{u}}
\newcommand{\bx}{\boldsymbol{x}}
\newcommand{\bU}{\boldsymbol{U}}
\newcommand{\bK}{\boldsymbol{K}}
\newcommand{\bOmega}{\boldsymbol{\Omega}}
\newcommand{\bldf}{\boldsymbol{f}}
\newcommand{\bS}{\boldsymbol{S}}
\newcommand{\bE}{\boldsymbol{E}}
\newcommand{\btau}{\boldsymbol{\tau}}
\newcommand{\bGamma}{\boldsymbol{\Gamma}}
\newcommand{\bsigma}{\boldsymbol{\sigma}}

The addition of colloidal particles to a fluid adds new stresses to the fluid arising from long-range hydrodynamic interactions between particles. Neglecting Brownian terms, the effective stress in the fluid can be written $\bsigma_e = \bsigma_f + \bsigma_h$, where $\bsigma_f = -p \boldsymbol{I} + \btau$ is the usual incompressible stress tensor for the fluid and $\bsigma_h$, sometimes called the hydrodynamic stress, comes from the addition of particles in the fluid \cite{Bossis_Brady_Rheology_1989}. Note we use $\btau = \eta \bE$, where $\bE$ is the rate of strain tensor. The added stress on the fluid will cause an effective increase in the viscosity of the suspension. Here, we look at systems driven by a simple shear that induces a background rate of strain
\begin{equation}
  \bE_\infty = \frac{1}{2} \begin{bmatrix}
    0                  & \dot{\vec{\gamma}} & 0 \\
    \dot{\vec{\gamma}} & 0                  & 0 \\
    0                  & 0                  & 0
  \end{bmatrix},
\end{equation}
where $\dot{\vec{\gamma}}$ is the shear rate. The resulting viscosity from each contribution to the stress tensor can be calculated as
\begin{equation}
  \label{eq:effective_viscosity}
  \eta_i = \frac{\bsigma_i : \bE_\infty}{\bE_\infty : \bE_\infty},
\end{equation}
where $i \in \left\{ e, f, h\right\}$ indicates which stress tensor is  being used to compute the viscosity. The viscosity of the fluid provides a natural scale, so we use the relative viscosity when comparing results,
\begin{equation}
  \eta_r = \frac{\eta_e}{\eta_f} = 1 + \frac{\eta_h}{\eta_f}.
\end{equation}

\begin{figure}
  \centering
  \includegraphics[width=0.99\linewidth]{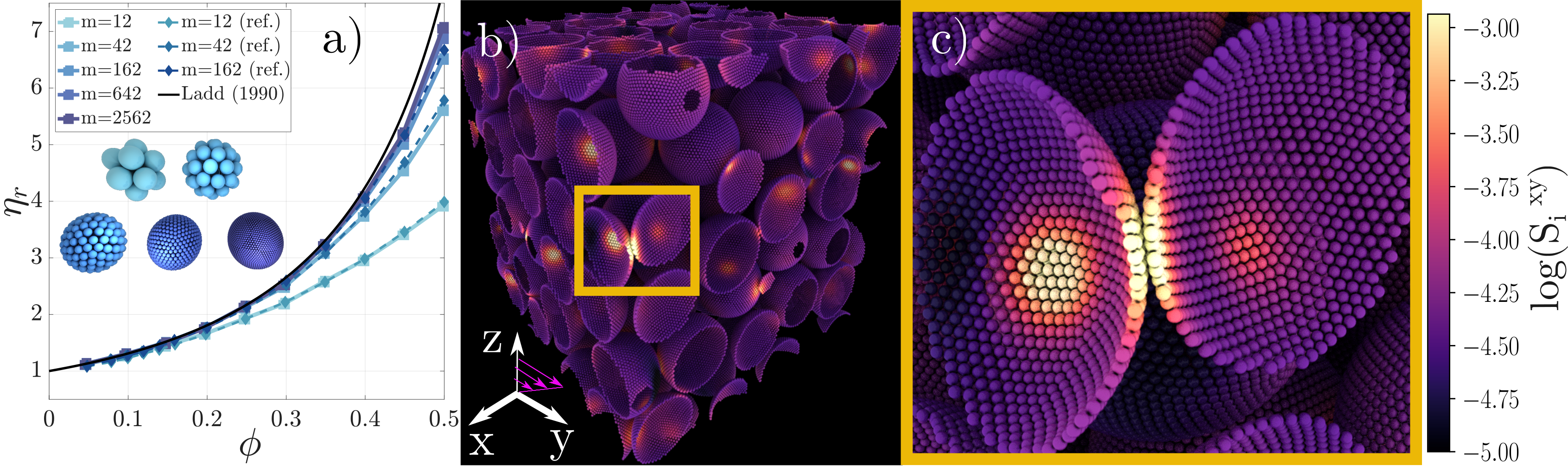}
  \caption{a) Relative viscosity $\eta_r$ of the sheared suspension for increasing packing fraction $\phi$ and different multiblob resolutions $m$. Reference data (dashed lines) is from Wang \textit{et al.} \cite{WangFioreSwan_ShearingRheology_2019} for m=12, 42, and 162. The semi-analytical reference curve (solid black line) is from equations (3.23) and (3.25) from Ladd \cite{Ladd_1990}. The inset spheres show the discretized multiblob spheres where the color corresponds to the resolution of the sphere (see legend). b) 3D rendering of one sample of the periodic suspension of spheres. The multiblobs are colored by the log of blob-wise stress. The axis in the lower left indicates the direction of shear. c) is the inset of figure b) showing the near-contact of two multiblob spheres. Blobs that appear missing are periodically wrapped to the opposite edge of the cube.}
  \label{fig:rheology_shear}
\end{figure}

Although \libM does not include stresslets in any of the solvers, it is still possible to study rheological properties. We demonstrate this by calculating the viscosity of a periodic suspension of colloids in a simple shear flow. Here, hydrodynamic contributions to the viscosity are calculated by discretizing a sphere representing a colloidal particle with many smaller blobs and using the rigid multiblob method \cite{Balboa2017}; see the insets on \cref{fig:rheology_shear}a) for an illustration of this discretization. The rigid multiblob method utilizes a geometric constraint matrix $\bK$ which that is used to map the velocity of the rigid body $\bU$ to surface velocities $\boldsymbol{u}$ on the individual blobs while maintaining rigidity as $\boldsymbol{u} = \bK \bU$. The transpose $\bK^T$ is also used to map forces on blobs $\bldf$ to the net force/torque on the body. If $\bldf$ is known, the stresslet on the $i-$th rigid body can be computed by summing over blobs $\alpha$ as \cite{WangFioreSwan_ShearingRheology_2019}
\begin{equation}
  \label{eq:discrete_stresslet}
  \bS^i = \frac{1}{2} \sum_{\alpha \in i} \left[ (\bx_\alpha - \bx_i) \bldf_\alpha^T + \bldf_\alpha(\bx_\alpha - \bx_i)^T \right].
\end{equation}
This can be seen as a discretized version of Equation 3 from Bossis \textit{et al.} \cite{Bossis_Brady_Rheology_1989} where $\bldf$ acts as the traction over the surface of the rigid body \cite{Delmotte_Usabiaga_multiblob_review_2025}. The velocity induced by the shear flow on each blob $\alpha$ in rigid body $i$ is $\bu_s = \bE_\infty (\bx_\alpha - \bx_i)$. Since $\vec{u_s}$ can be calculated directly, a resistance problem can be solved for $\bldf$. With the rigid constraints, the system is
\begin{equation}
  \label{eq:rigid_multiblob_system}
  \begin{bmatrix}
    \grandm & -\bK \\
    -\bK^T  & 0
  \end{bmatrix} \begin{bmatrix}
    \bldf \\ \bU
  \end{bmatrix} = \begin{bmatrix}
    \bu_s \\ 0
  \end{bmatrix},
\end{equation}
which is a modified version of Equation 16 from Wang \textit{et al.}\cite{WangFioreSwan_ShearingRheology_2019}. We solve this system using GMRES preconditioned with a diagonal approximation for $\grandm$ to a tolerance of $10^{-4}$. Once $\bldf$ is obtained, we can calculate the bulk hydrodynamic stress using \cref{eq:discrete_stresslet} as
\begin{equation}
  \bS_h = \sum_i \bS^i
\end{equation}
by summing over all rigid particles $i$. Neglecting the isotropic part of $\bsigma_h$ since it vanishes in the contraction with $\bE_\infty$, we can approximate $\bsigma_h \approx \bS_h$ and use \cref{eq:effective_viscosity} to compute the effective viscosity of the suspension.

To use the method described above, we first generate three samples of random periodic sphere packings at volume packing fractions $\phi$ ranging from 0.05 to 0.5 using code from Skoge \textit{et al.} \cite{Skoge_Donev_Torquato_sphere_packing_2006} in a periodic unit cell. The number of rigid spheres in the unit cell varies from 11 to 119. From these configurations, the average viscosity and standard error are computed using \libM's triply periodic \PSE solver across the range of packing fractions and for increasing resolution of the multiblob discretization. The results for viscosity are shown in \cref{fig:rheology_shear}a), and we note that error bars are smaller than the marker size. As the multiblob resolution increases, our simulation agrees closely with the semi-analytical model from Ladd \cite{Ladd_1990}. We note that a more accurate solution with fewer multiblobs can be obtained by adding a correction term that accounts for discretization error \cite{Moreno-Chaparro_Usabiaga_slipBCs_2025}. A depiction of the blob-wise stress for a sample of the largest simulations that use $2562$ blobs per colloid at a packing fraction of $\phi=0.5$ is shown in \cref{fig:rheology_shear}b) and c). For timing results for this simulation, see \cref{sec:timings}.

\subsection{Multi-physics (DPStokes)}
In this example, we study the fluid motion induced by traveling-wave electroosmosis (TWEO) over an array of microelectrodes actuated with phase-shifted AC signals, as depicted in \cref{fig:scheme}, in a slit channel filled with an electrolyte (salt water). This system recreates the experimental setup of P. García-Sánchez \textit{et al.} \cite{garcia-sanchez_experiments_2006}, where electroosmotic slip at the electrode surfaces generates net fluid transport in a confined microchannel. We introduce this surface velocity as an effective boundary condition and propagate it into the bulk. We compare the resulting bulk flow velocity to experimental measurements.
\begin{figure}[ht]
  \centering
  \begin{tikzpicture}
    \definecolor{liquid}{HTML}{6BD5FF}
    \definecolor{wall}{HTML}{005475}

    \definecolor{electrode}{HTML}{FFC801}
    \definecolor{pushers}{HTML}{FFE68A}

    \centering
    \def\circleRadius{0.2}
    \def\circleSpacing{0.8}
    \def\rectWidth{0.8}
    \def\rectHeight{0.2}
    \def\rectSpacing{0.8}

    \foreach \x in {0,...,9} {
        \foreach \y in {0,...,4} {
            \draw[fill=liquid] (\circleSpacing*\x, \circleSpacing*\y) circle (\circleRadius);
          }
      }

    \foreach \y in {0, ..., 10} {
        \draw[fill=wall] (\circleSpacing*10, 0.5*\circleSpacing*\y-0.5*\circleSpacing-\circleRadius) circle (\circleRadius);
      }

    \draw[->, thick, dashed] (-\rectSpacing, 1.5*\circleSpacing) -- ++(-1.5*\rectSpacing, 0);
    \draw[->, thick, dashed] (-\rectSpacing, 2.5*\circleSpacing) -- ++(-1.5*\rectSpacing, 0);
    \foreach \x [evaluate=\x as \index using int(\x+1)] in {0,...,4} {
        \coordinate (R) at (\rectSpacing*\x + \rectWidth*\x, -\circleSpacing);
        \draw[fill=electrode]
        (R) rectangle
        ++(\rectWidth, -\rectHeight);

        \path (R) ++(\rectWidth/2, -\rectHeight) coordinate (C);
        \pgfmathtruncatemacro{\angle}{mod(\index+2,4)*90}
        \draw[->, thick] (C) -- ++(0, -\circleRadius) node[below] {\large \angle$^o$};
        \draw[fill=pushers] (\rectSpacing*\x + \rectWidth*\x + 1*\circleRadius,-\circleSpacing + \circleRadius) circle (\circleRadius);
        \draw[fill=pushers] (\rectSpacing*\x + \rectWidth*\x + 3*\circleRadius,-\circleSpacing + \circleRadius) circle (\circleRadius);
      }

    \draw[thick] (-\circleSpacing, -\circleSpacing) -- (\circleSpacing*11, -\circleSpacing);
    \draw[->, thick] (-1.1*\circleSpacing, -\circleSpacing) -- ++(-0.5*\rectWidth, 0) node[left] {\large $v = 0$};
    \draw[thick] (-\circleSpacing, 4.5*\circleSpacing) -- (\circleSpacing*11, 4.5*\circleSpacing);
    \draw[->, thick] (-1.1*\circleSpacing, 4.5*\circleSpacing) -- ++(-0.5*\rectWidth, 0) node[left] {\large $v = 0$};

    \begin{scope}[shift={(-\circleSpacing, -\circleSpacing)}]
      \draw[->, thick] (-3,0) -- (-3,0.8*\circleSpacing) node[above] {$z$};
      \draw[->, thick] (-3,0) -- (-3+0.8*\circleSpacing,0) node[right] {$x$};
    \end{scope}

  \end{tikzpicture}
  \caption{Schematic representation of the system. Yellow rectangles at the base are the electrodes connected to an AC signal with a phase shift, indicated as angles in text below. Arrows on the left edge indicate that the system continues, repeating the 4-electrode array another 20 times. Blue circles are the tracer blobs where we evaluate the fluid velocity. Dark-blue blobs correspond to the vertical wall where we impose $\vec{v}^{\text{wall}} = 0$ and yellow blobs are where we impose the electroosmotic slip velocity $\vec{v}^{\text{electrode}} = \left<\vec{v}_{\text{slip}}\right>$, defined in \cref{eq:vslip}. The system is periodic in the y-direction (out of the page) and confined in the z-direction with two non-slip walls. The system resembles the experimental setup shown in Figure 1b from García-Sánchez \textit{et al.}\cite{garcia-sanchez_experiments_2006}, except discretized with blobs as described in-text.}
  \label{fig:scheme}
\end{figure}

\subsubsection{Methodology}
The density of free charges on an electrolyte fluid only differs from neutrality in the so-called electric double layer (EDL), which extends tens of nanometers away from the electrode's surface. The rest of the fluid can be treated as electro-neutral with the electric potential satisfying the Laplace equation,
\begin{equation}
  \nabla^2 \phi = 0.
  \label{eq:Laplace}
\end{equation}
Note that $\phi({\bf r})$ is a phasor representing the real, time-dependent potential $\mathrm{Re}\left[\phi({\bf r}) \,\exp[i\omega t]\right]$ and should be treated as complex valued.

As the EDL width, also called the Debye length, $\lambda_D$ is thin compared with the electrode width, $L$, it is possible to treat this thin region as an effective boundary condition (BC) for the potential $\phi$. This requires solving (or modeling) the inner domain to determine the outer potential at the EDL-bulk interface. We follow Ramos \textit{et al.}  \cite{Castellanos2002-3} who derived the EDL potential as a capacitor which gives,
\begin{equation}
  \sigma \frac{\partial \phi}{\partial z} = \frac{1}{Z_{DL}}(\phi - V_j),
  \label{eq:CC_DL}
\end{equation}
which is imposed at the surface of the electrode. Here, $\sigma$ is the conductivity of the medium and $Z_{DL}$ is the impedance associated with the EDL.
This BC would correspond to the EDL-bulk interface, but the thin-layer approximation ($\lambda_D/L<<1$) neglects the nanometric width of the EDL, allowing us to place the BC on the surface of the electrode. We assume that the EDL impedance is that of a perfect capacitor $Z_{DL} = (i\,\omega C)^{-1}$ with capacitance $C = \varepsilon/\lambda_D$. The characteristic EDL width is then given by
\begin{equation}
  \lambda_D = \left(\frac{\varepsilon k_BT}{n^0 q^2} \right)^{1/2}
\end{equation}
where $\varepsilon = \varepsilon_r\varepsilon_0$ is the absolute dielectric permittivity of the medium.
On the purely insulating walls (top and sides), the normal electric field is zero.
In the charged, thin layer formed over the patterned electrode surfaces where $\partial_x \phi \neq 0$, a net (oscillatory) tangential electric field forms and pushes ions to generate a flow. As initially shown by Ramos \textit{et al.} \cite{Castellanos2002-3},
\begin{equation}
  \left<\vec{v}_{\text{slip}}\right> = - \frac{\varepsilon}{4\eta} \Lambda \frac{\partial|\phi - V_j|^2}{\partial x}
  \label{eq:vslip}
\end{equation}
where $V_j$ is the (complex) potential connected to the electrode and $\Lambda\in[0,1]$ is a factor that determines the potential drop between the Stern and the compact layers \cite{Yang_Johnson_Wu_SternLayer2019}.

Using a finite difference Poisson solver, we solve for the electric potential and evaluate the slip velocity from \cref{eq:vslip} on a grid. We sample the slip velocity at the surface of the electrodes and interpolate from the mesh to obtain $\vec{v}^{\text{electrode}}$, which is discussed below. For this operation, we used the \texttt{spreadinterp} Python library developed by some of us \cite{pelaez2025spreadinterp}.

The \DPStokes solver in \libM can model a slit channel in which the $x$ and $y$ directions are periodic and the top and bottom of the domains are confined with no-slip walls. To model the system discussed above, we include extra confinement in $x$ and a slip boundary condition on the bottom surface. We do this by augmenting the domain with additional blobs to enforce constraints on the velocity using the immersed boundary method (IBM) \cite{IBM_PeskinReview}. The geometry is artificially modified by adding a no-slip wall created out of blobs (shown in dark blue in \cref{fig:scheme}) to confine the domain in the $x$ direction, modeling the end of the device. To include a slip velocity on the bottom surface, we include extra blobs just above the electrodes (shown in yellow in \cref{fig:scheme}). Similar to how some of us modeled electrokinetics in quadrupolar electrode systems \cite{dejong2025}, the extra blobs allow us to specify a boundary velocity $\vec{v}^\text{b}$ at each blob. Assuming that the only net force acting on the fluid is the one exerted by these blobs, we have
\begin{equation}
  \label{eq:electrode_resistance}
  \vec{v}^\text{b} = \tens{M}\vec{F}^\text{b},
\end{equation}
where $\vec{v}^b = \left[\vec{v}^{\text{wall}}, \vec{v}^{\text{electrode}}\right]^T$. We set $\vec{v}^{\text{wall}} = 0$ and $\vec{v}^{\text{electrode}}$ according to the slip velocity in \cref{eq:vslip}. This resistance problem can be solved for $\vec{F}^\text{b}$ using GMRES, similar to \cref{eq:rigid_multiblob_system}. We can then use $\vec{F}^b$ to calculate the fluid velocity at another point in the fluid by placing a tracer blob (light blue circles in \cref{fig:scheme}) and applying $\vec{F}^\text{b}$ to all boundary blobs and zero to all tracer blobs, e.g.
\begin{equation}
  \label{eq:electrode_velocity_field}
  \begin{bmatrix}\vec{v}^b \\ \vec{v}^{t} \end{bmatrix} = \tens{M}^{\text{tracer}} \begin{bmatrix}\vec{F}^b \\ 0 \end{bmatrix},
\end{equation} where $\tens{M}^{\text{tracer}}$ is calculated including the position of the tracer blobs, and is therefore larger than the mobility matrix in \cref{eq:electrode_resistance}, and $\vec{v}^t$ is the previously unknown velocity of the tracer particles. Unlike \cref{eq:rigid_multiblob_system}, this system has no additional constraints and thus the tracer particles do not interact with each other. Instead, the resulting velocity of each tracer blob represents a local average of the fluid velocity at that point and can be used to visualize velocity fields of the fluid. Notably, any solver from \libM is compatible with this procedure.

\begin{figure}[ht]
  \centering
  \includegraphics[width=\linewidth]{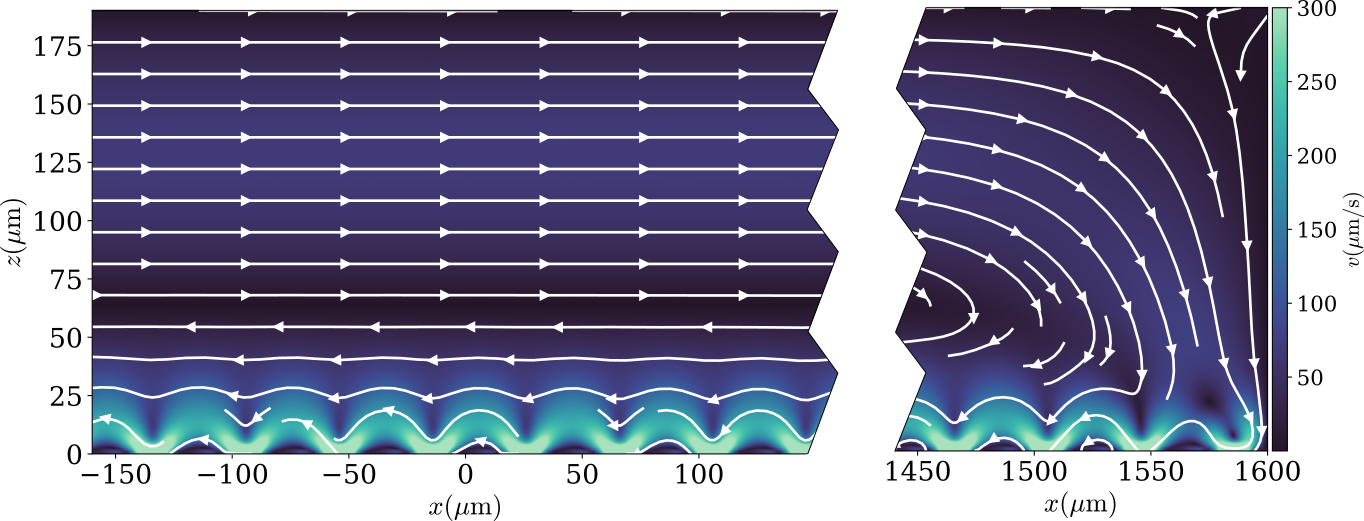}
  \caption{Streamlines of a 2D slice of TWEO, with the periodic direction coming out of the page. The colormap shows the velocity magnitude. A large portion of the domain is omitted for brevity, shown by the vertical white slash. Left: velocity field in the middle of the channel. Right: flow next to the wall placed at $x=1600\mu\text{m}$. The flow patterns observed here resemble those reported in the experimental results of Figure 4 from García-Sánchez \textit{et al.} \cite{garcia-sanchez_experiments_2006}.}
  \label{fig:streamPlot}
\end{figure}

The simulation domain was chosen to closely reproduce the geometry of the experimental device. It has dimensions $L_x = 3200 \mu\text{m}$, $L_y = 12\mu\text{m}$, and $L_z = 190\mu m$. The electrode width is $20 \mu\text{m}$, as is the electrode spacing. We use viscosity $\eta =10^{-3}\text{Pa}\cdot\text{s}$, density $\rho = 10^3\text{kg}/\text{m}^3$, and dielectric permittivity $\varepsilon = 80\varepsilon_0$. The electrostatic parameters include a Debye length $\lambda_D \approx 10\text{nm}$, an applied AC frequency $f = 2\text{kHz}$ and a potential drop $\Lambda = 0.03$. The IBM discretization uses blobs of hydrodynamic radius $a = 1\mu\text{m}$, Gaussian kernel width $4a$, and a cutoff radius of $12a$. The applied voltage at electrode $j$ is defined as $V_j = \frac{V_{pp}}{2}\, e^{\,(j\Delta\phi)i}$
where $V_{pp} =$ 6V is the applied peak-to-peak voltage and $\Delta\phi = \pi/2$ is the phase shift between neighboring electrodes (see \cref{fig:scheme}).

Streamlines of the resulting velocity field are shown in \cref{fig:streamPlot}. Our simulations successfully capture the main characteristics of the flow generated by TWEO in the experimental setup of García-Sánchez \textit{et al.}\cite{garcia-sanchez_experiments_2006}. The left portion of \cref{fig:streamPlot} shows the streamlines of the velocity field far from the vertical wall. Close to the electrode region, we observe strong oscillations due to the spacing of the electrodes, while farther from the surface the flow becomes smoother. Notably, the reversal of the flow direction at a certain height, due to backflow, is visible and consistent with the experimental observations. The right half of \cref{fig:streamPlot} shows the velocity field close to the vertical wall at $x=1600\mu\text{m}$ where the recirculation occurs.

\begin{figure}[h]
  \centering
  \includegraphics[width=0.7\textwidth]{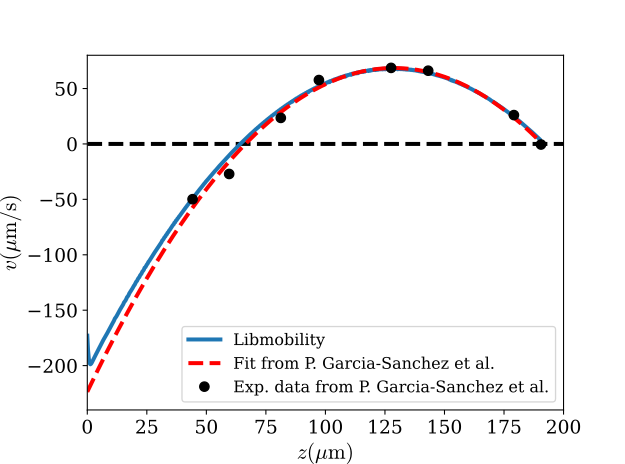}
  \caption{Horizontal velocity profile of TWEO as a function of height averaged along the interior of the electrode device for $V_{pp} = 6$V and $f=2$kHz. Black dots are experimental measurements and the dashed red line is the fit to Couette-Poisselle flow, both from García-Sánchez \textit{et al.}\cite{garcia-sanchez_experiments_2006}. The blue line is the horizontally averaged velocity from our numerical simulations.}
  \label{fig:profile}
\end{figure}

\Cref{fig:profile} presents the horizontal velocity profile as a function of height, averaged along the electrode array. The profile captures the smooth transition from the electrode-induced slip to the bulk flow, with increasing agreement as the influence of electrode size diminishes. Our numerical results show excellent agreement with both the experimental data and the Couette–Poiseuille empirical fit reported in the original work.

\subsection{Benchmarks}
\label{sec:timings}
In this section, we present benchmarks for the computation time required to compute deterministic particle displacements (the \texttt{Mdot} operation) for randomly placed particles, as well as timings for the simulations in sections \ref{sec:diffusion} and \ref{sec:rheology}. For the benchmarks, using randomized positions allows for overlapping particles, but this does not affect the runtime of \texttt{Mdot}. Particles that overlap boundaries can, however, affect the convergence of the Lanczos algorithm when computing \texttt{sqrtMdotW}, but this is not studied here. Benchmarks cover multiple geometries and boundary conditions, utilizing different \libM solvers.

The cost of computing the stochastic terms is solver-dependent. For example, \PSE is able to produce the fluctuating term alongside the deterministic one cheaply \cite{SpectralSD, ForceCoupling_Fluctuations}, and the thermal drift term can be neglected due to the triply periodic geometry. On the other hand, \DPStokes does not offer a specialized way to compute fluctuations or thermal drift, in which case \libM defaults to using the Lanczos algorithm for the former and RFD for the latter. Both algorithms are based on repeated application of the \texttt{Mdot} operation, and convergence of the Lanczos algorithm is dependent on the conditioning of the system. As such, timing information on the \texttt{Mdot} operation serves as a good overview of the performance of the library.

\begin{figure}
  \centering
  \begin{tikzpicture}
    \newcommand{\thickness}{1.0pt}
    \def\solverlist{
    1/DPStokes/DPStokes,
    2/PSE/PSE (non-Ewald),
    3/NBody/NBody,
    3/NBodyWall/NBody (bottom wall),
    4/{DPStokes_slit20}/DPStokes slit ($H=20a$),
    5/{PSE_split_custom}/PSE (Ewald)
    }
    \begin{groupplot}[
        group style={
            group size=2 by 1,
            y descriptions at=edge left,
            horizontal sep=0pt,
          },
        xmode=log,
        ymode=log,
        width=0.48\linewidth,
        cycle list={
            {red, line width=\thickness, mark=square*, mark size=3pt},
            {blue, line width=\thickness, mark=triangle*, mark size=3pt},
            {green!70!black, line width=\thickness, mark=diamond*, mark size=3pt},
            {orange, line width=\thickness, mark=pentagon*, mark size=3pt},
            {violet, line width=\thickness, mark=star, mark size=3pt},
            {cyan, line width=\thickness, mark=+, mark size=4pt},
            {magenta, line width=\thickness, mark=x, mark size=4pt},
            {brown, line width=\thickness, mark=asterisk, mark size=4pt},
            {olive, line width=\thickness, mark=oplus, mark size=3pt},
          },
        axis line style={very thick},
        tick style={very thick},
        minor tick style={thin},
        label style={font=\large},
        tick label style={font=\normalsize},
        xlabel={Number of particles},
        ymin=1e-5,
        ymax=100,
        legend style={at={(1.0,1.3)},anchor=north},
        legend columns=4,
        legend cell align=left,
        grid=major,
        major grid style={line width=.1pt,draw=gray!30}
      ]
      \nextgroupplot[ylabel=\texttt{Mdot} Time (s)]
      \foreach \idx/\solver/\name in \solverlist {
        \addplot table[
            x=nparticles,
            y={time_deterministic},
            col sep=comma
          ]{\solver_1.00e-01_False.csv};
        \addlegendentryexpanded{{\name}}%
      }
      \node[anchor=north east, font=\Large] at (axis cs:10000000, 1e-3) {$\rho=0.1a^{-3}$};
      \addplot[domain=100000:10000000, samples=10, ultra thick, black] {1e-8*x};
      \addlegendentry{$N$}
      \addplot[domain=100000:5000000, samples=10, ultra thick, black, dashed] {4e-12*x^2};
      \addlegendentry{$N^2$}
      \nextgroupplot[]
      \foreach \idx/\solver/\name in \solverlist {
        \addplot table[
            x=nparticles,
            y={time_deterministic},
            col sep=comma
          ]{\solver_1.00e-03_False.csv};
      }
      \node[anchor=north east, font=\Large] at (axis cs:10000000, 1e-3) {$\rho=0.001a^{-3}$};
      \addplot[domain=100000:10000000, samples=10, ultra thick, black] {3e-8*x};
      \addplot[domain=100000:5000000, samples=10, ultra thick, black, dashed] {4e-12*x^2};
    \end{groupplot}
  \end{tikzpicture}
  \caption{Comparison of solver timings at two different number densities, defined as $\rho=N/V$, with $V=L_xL_yL_z$ in units of the hydrodynamic radius, $a$. Particles are placed randomly in a domain that is cubic, unless specified otherwise. Only linear displacements (monopoles) are computed. The line \DPStokes uses the solver in a cubic domain ($L_x=L_y=L_z$), while the line \texttt{DPStokes slit} represents a domain with $L_x=L_y$ and $L_z=20a$ to simulate a slit channel for which this algorithm was designed and has superior performance. The \texttt{PSE (non-Ewald)} curve corresponds to a non-Ewald-split version of the algorithm, which is functionally equivalent to the traditional Force Coupling Method.\texttt{PSE (Ewald)} showcases the Ewald split version, with a splitting parameter $\xi=N^{1/3} / L$. Finally, the \texttt{NBody (bottom wall)} curve shows the \NBody solver configured with a no-slip wall at the bottom of the domain, which uses the slightly more expensive to compute RPY kernel with the wall correction from Swan and Brady \cite{StokesianDynamics_Wall}. Timings gathered using an RTX A6000 NVIDIA GPU.}
  \label{fig:timings_linear}
\end{figure}

Figure \ref{fig:timings_linear} presents timings for the different solvers applied to the same particle configurations. For a large number of particles, these curves highlight the algorithmic complexities of each solver as given by Table \ref{tab:solver_summary}. For systems with few particles, the performance curves tend to flatten which indicates the runtime is dominated by overhead costs that are due in large part to an under-utilization of the GPU’s computational capacity. Most of these algorithms parallelize the load by assigning threads to either particles, discrete pieces of the domain, or both. As such, `small' systems might be sub-optimally load-balanced. Modern GPUs present around ten thousand CUDA cores, which \emph{very roughly} translates into the scaling crossover occurring around ten thousand particles for purely particle-based solvers such as \texttt{NBody}. Finally, note that the benchmarks presented in this section do not include the any time to transfer inputs and/or outputs between the CPU and the GPU. Using \libM with user code that provides inputs on the CPU comes with the additional overhead of transferring the inputs and outputs to and from GPU memory, although this transfer is taken care of automatically by \libM. This overhead can often dominate the computation time in small systems, even if the library attempts to mitigate some of it by overlapping transfers and computation or using host-mapped memory. Using other GPU-first libraries (such as \texttt{cupy} or \texttt{pytorch}) to provide inputs and process outputs can nullify this overhead.

\begin{figure}
  \centering
  \includegraphics[width=0.9\linewidth]{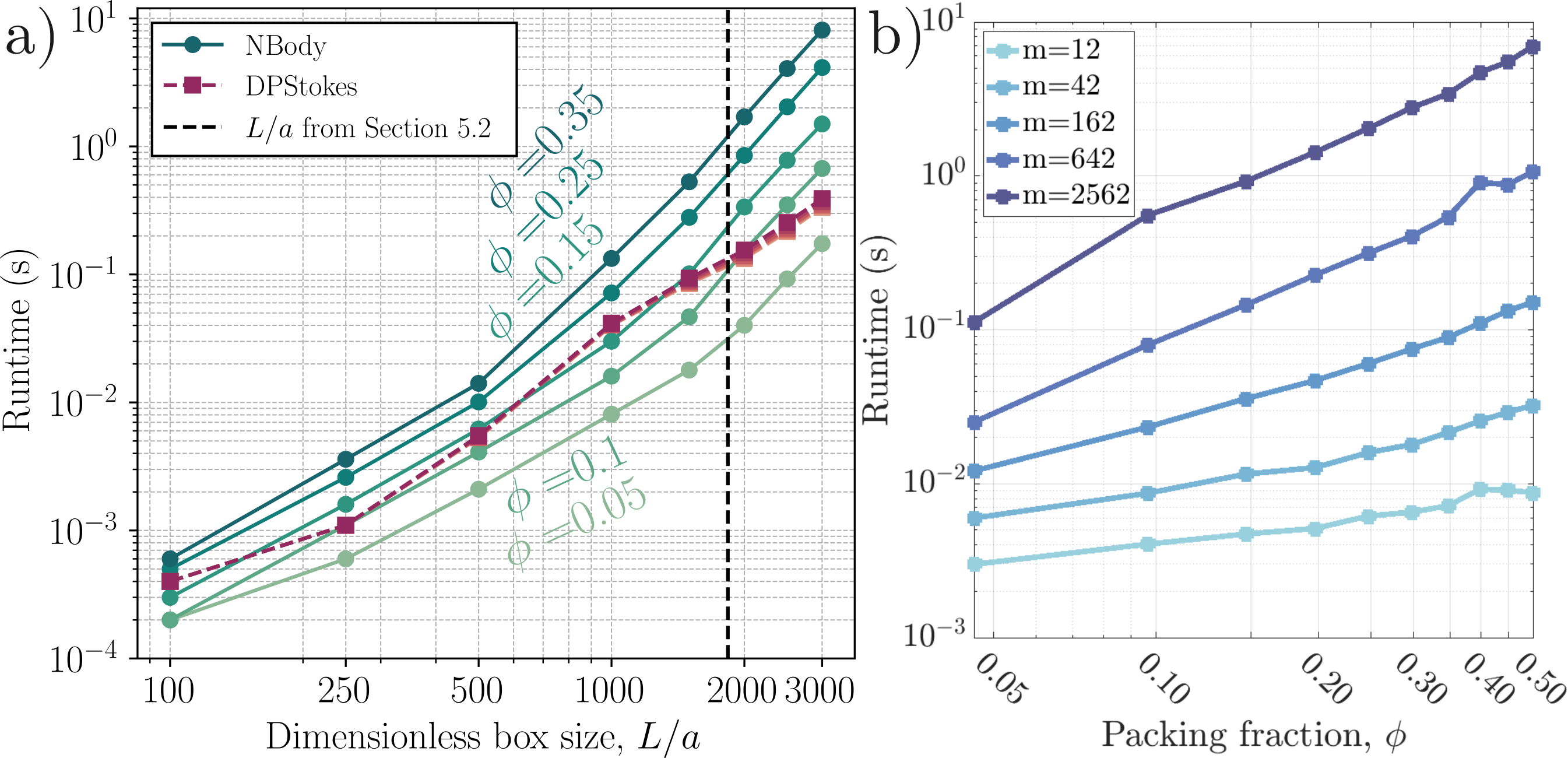}
  \caption{a) Comparing runtimes of one \texttt{Mdot} between \NBody and \DPStokes for the diffusing particle case in \cref{sec:diffusion} for different box sizes $L$ and (in-plane) packing fraction $\phi$ in log-log scale. The curves for \DPStokes for varying $\phi$ are largely collapsed onto each other. The vertical dashed line indicates the domain size used in \cref{sec:diffusion}. Timings gathered using an NVIDIA A40 GPU. b) Computational time for solving the system in \cref{eq:rigid_multiblob_system} for the cases shown in \cref{fig:rheology_shear}a) using the \PSE solver versus the (volume) packing fraction in log-log scale. Colormap indicates increasing resolution of multiblob discretizations, denoted $m$ in the legend. The domain size is fixed, so increasing the packing fraction comes from increasing the number of spheres within the domain. Timings gathered on a NVIDIA RTX 4070 GPU.}
  \label{fig:experiment_timings}
\end{figure}

We also present timings for simulations in \cref{sec:diffusion} and \cref{sec:rheology}. \Cref{fig:experiment_timings}a) shows the cost for computing one \texttt{Mdot} for the \NBody and \DPStokes solvers for various domain sizes at different packing fractions. As these solvers both support the \texttt{single\_wall} geometry, with the only difference being open vs. periodic conditions in the $xy$-plane, a user is often free to choose between these solvers for an application. In this scenario, the best solver to use will usually be the one with the lowest computational cost. \Cref{sec:diffusion} presents a somewhat challenging case since the domain size is large, which is costly for \DPStokes, but the number of particles grows quickly with domain size for a fixed packing fraction, with a corresponding increase in the cost of the \NBody solver. The dominant cost of \DPStokes is the Stokes solve, so the solver's runtime is somewhat insensitive to the number of particles compared to the cost of solving the fluid equations on a larger grid. This can be seen in \cref{fig:experiment_timings} as all the \DPStokes curves for varying packing fraction ($0.05 \leq \phi \leq 0.35$) are largely on top of each other. In contrast, the cost of the \NBody solver is independent of domain size, but scales quickly with total number of particles. As such, \NBody is dramatically more expensive to use than \DPStokes at high packing fractions $(\phi \geq 0.15)$ but can be cheaper for low packing fractions, especially for small domains. These timings show the parameters of $\phi=0.11$ and $L/a \approx 1800$ considered in \cref{sec:diffusion} to be slightly more efficient when using \NBody on the NVIDIA A40 GPU used for those simulations, but we note that the exact location of the crossover is hardware dependent. In our experience, the performance of \DPStokes improves on graphics cards with higher memory bandwidth.

\Cref{fig:experiment_timings}b) shows times to solve the system in \cref{eq:rigid_multiblob_system} for increasing packing fraction and varying sphere discretizations; see \cref{sec:rheology} for more details. For a fixed $m$, we see an increase in runtime as packing fraction increases, as well as an increase in runtime for larger $m$. Both larger $\phi$ and $m$ increase the total number of blobs in the system, but a larger $\phi$ will also increase the total number of GMRES iterations required to solve the system.

\section{Conclusions}
\label{sec:conclusions}
In this work, we introduced \libM, a high-performance Python-based software library specifically designed to simulate hydrodynamic interactions in particulate systems at the Rotne-Prager-Yamakawa level, optimized for execution on NVIDIA GPU architectures. \libM addresses critical computational challenges associated with hydrodynamic modeling by providing efficient, ready-to-use solvers for complex particle-fluid dynamics under a unified interface.

The modular design of \libM facilitates flexibility, allowing integration and extension of new solvers to cover various geometries and boundary conditions. Currently, \libM includes solvers tailored for fully open, triply periodic, singly confined, and doubly periodic systems. Extensive validations performed through rigorous tests confirm that the implemented methods adhere closely to theoretical predictions, particularly respecting fluctuation-dissipation relationships and correctly capturing thermal drift effects.

Practical demonstrations highlight \libM's applicability in diverse scenarios. The library’s GPU acceleration enables simulations at scales and complexities previously considered infeasible with CPU-based implementations, significantly expanding the computational horizons available to researchers in computational fluid dynamics, biophysics, and materials science. Notably, the scalability and efficiency of \libM enabled simulations that are directly comparable in scale to experimental systems.

Future developments will focus on expanding solver functionalities, further optimizing performance, and broadening accessibility across additional computing platforms. We anticipate that \libM will become a valuable resource, enabling detailed hydrodynamic investigations and accelerating discoveries across scientific disciplines that rely on accurate fluid-particle modeling.

\section*{Acknowledgements}
We wholeheartedly thank Professor Aleksandar Donev for his initiative and support in kick-starting the development efforts for this work.

A.C. acknowledges the Institute of Materials Science (iMAT) of the Alliance Sorbonne Université for a PhD grant. This work was granted access to the HPC resources of the SACADO MeSU platform at Sorbonne Université. R.D-B, R.P.P. and P.D.S. acknowledge funding form HORIZON-EIC-2023-PATHFINDEROPEN(FASTCOMET), grant 101130615.

\section*{Author declarations}
The authors have no conflicts to disclose.

\section*{Data availability}
The data that supports the findings of this study are available within the article and the associated code repositories.

\bibliography{References}

\newpage

\end{document}